\documentclass[useAMS,usenatbib,longabstract]{mn2e}
\usepackage{graphicx,fleqn}
\usepackage{deluxetable}
\usepackage{longtable}

\def \apj{ApJ}
\def \apjl{ApJL}

\def\ltsim{\raisebox{-.5ex}{$\;\stackrel{<}{\sim}\;$}}
\def\gtsim{\raisebox{-.5ex}{$\;\stackrel{>}{\sim}\;$}}

\title[GRB 130831A central engine]{The central engine of GRB 130831A and the energy breakdown of a relativistic explosion.}
\author[]{M. De Pasquale $^{1,2,3}$, S. R. Oates$^{1,4}$, J. L. Racusin$^{5}$,  D. A. Kann$^{6}$, B. Zhang$^7$, 
\newauthor  A. Pozanenko$^{8,27}$, A. A. Volnova$^8$, A. Trotter$^9$, N. Frank$^9$, A. Cucchiara$^5$, E. Troja$^5$, 
\newauthor  B. Sbarufatti$^{10}$, N. R. Butler$^{11}$, S. Schulze$^{12,13}$, Z. Cano$^{14}$,  M. J. Page$^{1}$, 
\newauthor  A. J. Castro-Tirado$^{4,26}$,  J. Gorosabel$^{4, 15, 16, \dagger}$, A. Lien$^{5,17}$, O. Fox$^{18}$, O. Littlejohns$^{11}$, 
\newauthor   J. S. Bloom$^{18}$,  J. X. Prochaska$^{19}$, J. A. de Diego$^{20}$,  J. Gonzalez$^{20}$, M. G. Richer$^{21}$, 
\newauthor  C. Rom\'an-Z\'u\~{n}iga$^{21}$,  A. M. Watson$^{20}$,  N. Gehrels$^5$, H. Moseley$^5$, A. Kutyrev$^{5}$, S. Zane$^1$, 
\newauthor V. Hoette$^{22}$,  R. R. Russell$^{22}$,  V. Rumyantsev$^{23}$, E. Klunko$^{24}$, O. Burkhonov$^{25}$, 
\newauthor A. A. Breeveld$^1$,  D. E. Reichart$^{9}$, J.B. Haislip$^{9}$\\
$^{1}$Mullard Space Science Laboratory, University College London, Dorking, United Kingdom E-mail: m.depasquale@ucl.ac.uk\\
$^{2}$Istituto Astrofisica Spaziale Fisica Cosmica, Palermo, Italy\\
$^3$Istituto Euro Mediterraneo di Scienza e Tecnologia, Palermo, Italy\\
$^4$Istituto de Astrofisica de Andalucia (CSIC), Granada, Spain\\
$^5$Center for Research and Exploration in Space Science And Technology (CRESST) and NASA Goddard Space Flight Center, Greenbelt\\
MD 20771, USA\\
$^6$Th\"uringer Landessternwarte Tautenburg, Sternwarte 5, 07778 Tautenburg, Germany\\
$^7$Department of Physics and Astronomy, University of Nevada Las Vegas, Las Vegas, USA\\
$^8$Space Research Institute (IKI), Moscow, Russia\\
$^9$University of North Carolina at Chapel Hill, Chapel Hill, North Carolina, USA\\
$^{10}$Pennsylvania State  University, University Park, PA, USA\\
$^{11}$Arizona State University, Tempe, Arizona, USA\\
$^{12}$Instituto de Astrof\'isica, Facultad de F\'isica, Pontificia Universidad Cat\'olica de Chile, Vicu\~{n}a Mackenna 4860, 7820436 Macul, Santiago, Chile\\ 
$^{13}$Millennium Institute of Astrophysics, Vicu\~{n}a Mackenna 4860, 7820436 Macul, Santiago, Chile\\
$^{14}$Centre for Astrophysics and Cosmology, Science Institute, University of Iceland, 107 Reykjavik, Iceland.\\
$^{15}$Unidad Asociada Grupo Ciencia Planetarias UPV/EHU-IAA/CSIC, Departamento de F\'{\i}sica Aplicada I, E.T.S. Ingenier\'{\i}a, Universidad del Pa\'{\i}s \\
Vasco UPV/EHU, Alameda de Urquijo s/n, E-48013 Bilbao, Spain\\
$^{16}$Ikerbasque, Basque Foundation for Science, Alameda de Urquijo 36-5, E-48008 Bilbao, Spain\\
$\dagger$ Deceased\\
$^{17}$Department of Physics, University of Maryland, Baltimore County, Baltimore, MD 21250, USA\\
$^{18}$University of California Berkeley, USA\\
$^{19}$University of California Santa Cruz, USA\\
$^{20}$Instituto de Astronom\'ia, Universidad Nacional Aut\'onoma de M\'exico, Apartado Postal 70-264 04510 M\'exico, DF, Mexico\\
$^{21}$Instituto de Astronom\'ia, Universidad Nacional Autonoma de M\'exico, Apartado Postal 106, 22800 Ensenada, Baja California, Mexico\\
$^{22}$The University of Chicago, IL, USA\\
$^{23}$Crimean Astrophysical Observatory, 98409, pgt. Nauchny, Crimea\\
$^{24}$Institute of Solar-Terrestrial Physics, Russian Academy of Sciences, 664033, p/o box 291; Lermontov st., 126a, Irkutsk, Russia\\
$^{25}$Ulugh Beg Astronomical Institute, 100052, 33 Astronomicheskaya str., Tashkent, Uzbekistan\\
$^{26}$Unidad Asociada Departamento de Ingenieria de Sistemas y Automatica, Universidad de Malaga, Spain\\
$^{27}$National Research Nuclear University MEPhI (Moscow Engineering Physics Institute), 115409 Moscow, Russia}

\begin{document}

\date{Accepted ... Received ...}

\maketitle

\label{firstpage}

\clearpage

\begin{abstract}
Gamma-ray bursts (GRBs) are the most luminous explosions in the universe, yet the nature and physical properties of their energy sources are far from understood.Very important clues, however, can be inferred by studying the afterglows of these events. We present optical and X-ray observations of GRB 130831A obtained by {\it Swift}, {\it Chandra}, Skynet, RATIR, Maidanak, ISON, NOT, LT and GTC. This burst shows a steep drop in the X-ray light-curve at $\simeq 10^5$ s after the trigger, with a power-law decay index of $\alpha \sim 6$. Such a rare behaviour cannot be explained by the standard forward shock (FS) model and indicates that the emission, up to the fast decay at $10^5$~s, must be of ``internal origin", produced by a dissipation process within an ultrarelativistic outflow. We propose that the source of such an outflow, which must produce the X-ray flux for $\simeq 1$ day in the cosmological rest frame, is a newly born magnetar or black hole.
After the drop, the faint X-ray afterglow continues with a much shallower decay. The optical emission, on the other hand, shows no break across the X-ray steep decrease, and the late-time decays of both the X-ray and optical are consistent. Using both the X-ray and optical data, we show that the emission after $\simeq 10^5$ s can be explained well by the FS model. We model our data to derive the kinetic energy of the ejecta and thus measure the efficiency of the central engine of a GRB with emission of internal origin visible for a long time. Furthermore, we break down the energy budget of this GRB into the prompt emission, the late internal dissipation, the kinetic energy of the relativistic ejecta, and compare it with the energy of the associated supernova, SN 2013fu.
\end{abstract}


\section{Introduction}

The study of gamma-ray bursts (GRBs) has received an exceptional boost thanks to the {\it Swift} mission (Gehrels et al. 2004), which has enabled rapid follow-up radio to X-ray observations of GRBs. However, despite a very large number of such fast follow-up observations performed by this spacecraft and by ground observatories, the characteristics of the ``central engine" that produces the GRB are still unclear. 
%This lack of knowledge is due to the fact that most information on GRBs is inferred from the ``afterglow" emission, which does not directly comes from the outflow but from
 The prevailing model (Woosley 1993; MacFadyen 1999, 2001; Thompson 2007) predicts the formation of a compact object, either a black hole or a magnetar, surrounded by an accretion disk. The compact object is the result of the core collapse of a very massive star or the final merger of two neutron stars (NS-NS)  or NS - black hole binary. Under the correct conditions of angular momentum and magnetic field, such a system would launch a collimated, ultra-relativistic jet (e.g. via the Blanford-Znajek mechanism; Blandford \& Znajek 1977). Within these jets, one or more ``internal dissipation" processes take place (for a review, see Zhang 2011), converting part of the kinetic and/or magnetic energy into radiation. A burst of gamma-rays will then be visible if the observer is placed within the opening angle of the outflow. 
 
Following the prompt gamma-ray emission, long-lived afterglow emission is detected in several energy bands (radio, optical, X-ray and probably high-energy gamma-rays). The consensus is that the afterglow radiation is emitted when ultra-relativistic ejecta interact with the circumburst medium, driving a forward shock (FS), which moves into the medium, and a reverse shock (RS), which propagates backwards through the ejecta. In particular, the emission due to the FS can in principle last indefinitely. A hallmark of RS and FS afterglow emission is that the flux density $F_{\nu}$ behaves as a power-law both in time and frequency, being described as $F_{\nu} \propto t^{-\alpha} \nu^{-\beta}$, where $t$ is the time from the GRB trigger (Kobayashi \& Zhang 2007) and $\nu$ is the frequency. \\
However, {\it Swift} observations have produced evidence of more complex phenomena during the afterglow phase, such as X-ray and optical flares (Falcone et al. 2006; Margutti et al 2011; Swenson et al. 2013), that cannot be attributed to the FS due to their fast temporal variability. For a few kiloseconds (ks) after the trigger, the afterglow flux often decays in a slow fashion (phase II of the canonical XRT light-curve, also known as ``plateau"; O ' Brien et al. 2006, Nousek et al. 2006; Racusin et al. 2009; Margutti et al. 2010) which cannot be understood if the fireball follows an adiabatic evolution; a process of energy injection (Zhang et al. 2006) into the fireball is usually invoked to explain such a feature. These observations strongly suggest that the GRB central engine is still active, and produces energy and/or a relativistic outflow. 

In a small subset of {\it Swift} GRB afterglows, observations have shown slow decline phases of the X-ray flux which terminate with an abrupt fall in the emission, with slopes $\alpha \gtsim 3-4$, sometimes approaching $\alpha \simeq 9-10$ (Troja et al. 2007, Liang et al. 2007, Lyons et al. 2010, Rowlinson et al. 2013; L\"u \& Zhang 2014). Again, the FS model does not predict such behaviour. Instead, such steep decay can be expected when the central engine is a newly born magnetar, which emits a very high luminosity outflow due to the spin-down process (Usov 1992, Zhang \& Meszaros 2001, Zhang \& Meszaros 2002, Dall'Osso et al. 2011), which in turn produces emission we can directly observe in the X-ray. A very energetic outflow could also be produced by a newly formed stellar mass black hole surrounded by an accretion disk. The electromagnetic luminosity is expected to fall rapidly after the time-scale $T_{\rm em}$ of the spin-down process, if the magnetar collapses into a black hole, or accretion onto the black hole stops. 

The outflow produced by the spin-down process should generate emission via the synchrotron process. This kind of emission seems to be produced mostly in the X-ray band, while it is not usually detected at lower frequencies such as the optical (see Troja et al. 2007, Zhang et al. 2009, Rowlinson et al. 2013; however see Cano et al. 2014 for GRB130215A), perhaps because the optical band is at or below the synchrotron self-absorption frequency (Zhang 2009; see also Shen \& Zhang 2009). At low frequencies, the dominant emission mechanism seems to be the standard FS, with its power-law decays and slow flux variations.
FS emission, however, is still expected to be present even in the X-ray band, and it should emerge once the X-ray emission from the outflow produced by the magnetar or the black hole drops. So far, this has been seen clearly in the case of the long GRB 070110 (Troja et al. 2007). In this event, the X-ray light-curve showed a plateau lasting for ~20 ks, after which the flux fell quickly with a slope of $\alpha \sim 9$, and then resumed a power-law decay with a much shallower slope of $\alpha \sim 1$. Such a late slope appears in most GRB afterglows and is likely produced by the FS. 

In this paper, we present the well-sampled X-ray, UV/Optical and NIR observations of the afterglow of GRB 130831A, and show that its behaviour can be interpreted as a superposition of FS emission and ``internal emission", the latter of which suddenly ceases at $\simeq 100$~ks.
This article is organized as follows. In Sect. 2, we present the observations taken by different instruments and observing facilities, and show how the data were reduced and analyzed. We recall the results of the observations taken by other groups on GRB130831A as well. In Sect. 3, we show and fit the resulting light-curves and spectral energy distributions of this GRB. In Sect. 4, we model the afterglow of GRB 130831A in the context of FS and internal emission models, we discuss the possible origins of the observed emission and the properties of the object that produced the explosion. Finally, we present our conclusions in Sect. 5. 
% The X-ray and optical light-curves of this GRB had typical slopes until 100 ks, when the X-ray showed a very steep decay with a decay slope of $\sim 6$. This very rapid decay cannot be attributed to FS emission. At about $\simeq 170$~ks, the decay rate changed again, giving way to a more ``standard" slope of $\alpha \sim 1.1$. The optical light-curve, which is richly sampled, shows no break simultaneous to that seen in the X-ray. We discuss the properties of  {GRB 130831A} and interpret its light-curves as activity of a ``central engine", either a black hole or a magnetar, responsible for the early X-ray emission, coupled with the standard FS mechanism, which powers the late X-ray and global optical emission. Furthermore, we present a prospect of the energetics associated with the early X-ray emission, the prompt $\gamma$-ray emission, the relativistic outflow, and the SN associated with GRB 130831A.\\
Throughout the paper, errors are expressed at the $1\sigma$ confidence level (CL) unless stated otherwise, and we assume a $\Lambda$CDM Cosmology with $H_0 = 70$ km $s^{-1}$ Mpc$^{-1}$, $\Omega_{\rm m}= 0.27$ and $\Omega_{\Lambda}=0.73$ (Jarosik et al. 2011).

\section{Observations and Analysis}

\subsection{X-ray data}\label{xrdata}

\subsubsection{{\it Swift}-BAT} The {\it Swift} Burst Alert Telescope (BAT;
Barthelmy et al. 2005) triggered on GRB 130831A at $T_0=$13:04:16.54 UT,
2013 August 31. The mask-weighted light-curve in the 15-350~keV energy range
(see Fig. 1) shows a main pulse with a fast-rise-and-exponential-decay (FRED)
shape. It starts at {\it T}$_0 $-2~s and peaks at {\it T}$_0 $+3 s. The major
pulse structure ends at {\it T}$_0$+12 s. However, there is some extended
emission that lasts until {\it T}$_0$+41 s, with two additional peaks at {\it T}$_0
$+20 s and {\it T}$_0$+32 s. {\it T}$_{90}$ (15-350 keV), the interval during which from 5\% to 95\% of the total emission is recorded, is $30.2 \pm 1.4$~s (error includes systematics).

The time-averaged spectrum between {\it T}$_0-$1.9 s to {\it T}$_0$+41.4 s can be fitted by a simple power-law model with a spectral index $\beta=0.93 \pm 0.03$ ($\chi^2 = 56.7$ for 57 degrees of freedom, $dof$) in the energy band 15-150 keV. This model gives a fluence of $(6.49 \pm 0.09) \times 10^{-6}$ erg cm$^{-2}$ in the same energy range. The 1-s peak spectrum can also be fitted by a simple power-law model with $\beta=0.70 \pm 0.05$ ($\chi^2/dof = 56.4/57$) in the energy band 15-150 keV, which gives a peak flux of $9.44 \pm 0.25 \times 10^{-7}$ erg cm$^{-2}$ s$^{-1}$ in the same band. The BAT analysis uses the event data from $T_0$-240 s to $T_0$+963 s. 

\begin{figure}
\begin{center}
%\epsscale{.80}
%\plottwo{GRB130831_optical.ps}{fluxcurve_final_new.ps}
\includegraphics[angle=0,width=8cm]{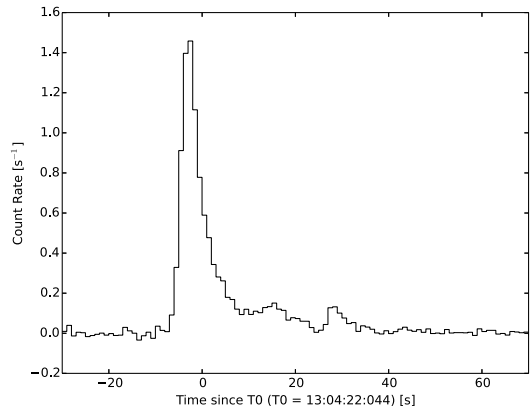}
\caption{{\em Swift} BAT light-curve of the prompt emission from GRB~130831A.} 
%{\it Konus-Wind} light-curve is taken from {$www.ioffe.ru/LEA/GRBs/GRB130831\_T47062$}}
\label{prompt}
\end{center}
\end{figure}

%\subsection{\it XRT and {\it Chandra} ACIS data}

%The {\it Swift} XRT started observations of  {GRB 130831A} 126 s after the the BAT trigger and detected a bright and uncatalogued source. The position of the source was refined following the technique by Goad et al. 2007 (A\&A, 476, 1401) and determined as RA = 23$^{0}$ 54m 29.91 s, Dec 29 25' 47.6'', with an uncertainty of 1.7''.

\subsubsection{{\it Swift}-XRT Observations}
After the BAT trigger, the {\it Swift} satellite promptly slewed to point the narrow field instruments at the source. The X-ray Telescope (XRT; Burrows et al. 2005) began observations of GRB 130831A~125.8 s after the trigger in the 0.3--10 keV energy band.
% A new uncatalogued source was detected at RA = 23$^{0}$ 54m 29.91 s, Dec 29 2x5' 47.6'', with an uncertainty of 1.7 arc seconds.
The XRT monitored the source until 2013 September 14, collecting 84.9 s
of data in Window Timing (WT) mode (11.6 s were taken while the spacecraft was
settling and the remaining 73.3~s pointing mode) and 59.7 ks in Photon
Counting (PC) mode. 

XRT data were processed using the HEASoft software packages\footnote{http://heasarc.nasa.gov/lheasoft/}, version 6.15 and version 20130313 of the XRT Calibration DataBase\footnote{www.swift.ac.uk/analysis/xrt/files/SWIFT-XRT-CALDB-09\_v17.pdf}, applying calibrations and standard filtering and screening criteria (for more details, see Evans et al. 2009).
WT data were extracted in an interval centred on the source, 20 pixels on each side, and the background estimated using intervals between 40 and 60 pixels from the source. The PC data were initially affected by pile-up, and corrected by
excluding the central core of 6 pixel radius. The remaining PC data were extracted
using a circular region with 30-pixel radius, and when the count rate
dropped below $10^{-2}$~counts~s$^{-1}$, within a 10 pixel radius. During the extraction of the light-curve, a minimum signal-to-noise ratio (S/N) criterion was applied for the re-binning, such that we required a S/N greater than 3 for all data points, with the exception of the last data point. The 90\% confidence level intervals were estimated following the Kraft et al. (1991) technique for low count rate sources.  The X-ray light-curve (Fig. 2), which shows the temporal evolution of the flux, was constructed once the spectral information was obtained (see Sect.s \ref{spectral analysis} and \ref{xraylc}, Tables \ref{tab:xrtspec} and \ref{tab: xrt time analysis}).

\subsubsection{{\it Chandra} observations}
After the very steep break, we requested and obtained two {\it Chandra}
Director's Discretionary Time (DDT) observations (PI: De
Pasquale). Meanwhile, {\it Swift} continued observing for several more days
detecting a dim X-ray afterglow, which suggested that the steep decay had
broken to a gentler slope. The {\it Chandra} observations of GRB
130831A took place on 2013 September 17 (T$_0$+16.6 days, or 1430 ks) and
2013 October 3 (T$_0$+33.1 days, or 2860 ks) with an exposure of 15 ks
each. The location of GRB~130831A was imaged on the ACIS S3 chip in 
both observations. The ACIS data were processed using 
{\sc ciao} version 4.6 and version 4.5.9 of the Calibration Database (Fruscione et al. 2006) using standard tools. We filtered the events for grades 0, 2, 3, 4, and 6 and energy from 0.5--8 keV. We extracted the source counts within a $1.5\farcs$ radius region centered on the best known position of the GRB, yielding 8 and 1 counts in the two observations, respectively. The first epoch yielded a detection with a significance of $5.4\sigma$ determined using the method described in Kraft et al. (1991). The second epoch was only $\sim 1\sigma$ above background, thus it should not be regarded as a real detection.

{\subsection{Optical observations}\label{OIR}\label{optical}

Here we describe how we collected, reduced and calibrated 
the ultraviolet, optical and infrared photometry of GRB~130831A. All magnitudes
have been calibrated to the Vega system. The full set of
photometric measurements is provided in an on-line table; a sample is shown in Table~\ref{Photometry}.

\subsubsection{UVOT}
{\it Swift}/UVOT (Roming et al. 2005) began observing the field of GRB 130831A 114 s after 
the trigger (Hagen et al., GCN Circular 15139) and started settled observations
191 s after the trigger, with a finding chart exposure in the $u$ band. The
afterglow was detected in all 7 UVOT filters.  Observations were taken in both
imaging and event modes. Before extracting count rates  from the event lists, the
astrometry was refined following the method described in Oates et
al. (2009). The source counts were extracted initially using a source region of
5" radius. When the count rate dropped to below 0.5 counts per second we then
used a source  region of 3" radius. In order to be consistent with the UVOT
calibration, these count  rates were then corrected to 5" using the curve of
growth contained in the calibration files.  Background counts were extracted
using a circular region of radius 20" from a blank area  of sky situated near
to the source position. The count rates were obtained from the event and image
lists using  the {\it Swift} tools {\sc uvotevtlc} and {\sc uvotsource},
respectively. They were converted to magnitudes using the UVOT
photometric zero points (Breeveld et al., 2011). The analysis pipeline used
version 20130118 of the UVOT Calibration Database.
%The UVOT data is provided 
%in Table \ref{uvoir}. The analysis pipeline used UVOT calibration 20130118.
%The optical afterglow of GRB 130831A was also detected by other facilities, including SKYNET (Trotter et al., 2013) in $B$, $V$, $R$, $I$ filters, RATIR (Butler et al. 2013) $g'$, $i'$, $Z$, $Y$, $J$, $H$ filters, the ISON network (Volnova et al. 2013a; Pozanenko et al. 2013a), Maidanak (Pozanenko et al. 2013b), the 2.6-m Shajn telescope of Crimean Astrophysical Observatory and with the 1.6-m telescope AZT-33IK of Sayan observatory with unfiltered exposures and $V$, $B$ $R$, and $I$ filters, Nordic Optical Telescope (NOT) and the Gran Telescopio Canarias (GTC) in $r'$ and $i'$ filters. 

\subsubsection{RATIR}
The Reionization And Transients Infra-Red camera (RATIR; Butler et al. 2012)
observed GRB~130831A over a period of seven hours, beginning 15.8 hours after
the {\it Swift} trigger, with follow-up observations on six nights over the
next month.  RATIR is mounted on the 1.5-meter Harold L. Johnson telescope of
the Observatorio Astron�omico Nacional on Sierra San Pedro M�artir in Baja
California (Mexico). This facility, which became fully operational in
December 2012, conducts autonomous observations (Watson et al. 2012; Klein et
al. 2012) of its targets in the six photometric bands  $r'$, $i'$, $z'$, $Y$,
$J$ and $H$ simultaneously. RATIR captured 80~s exposure frames in $r'i'$ and
67~s exposure frames in $z'YJH$ due to additional overhead. We
applied their standard image reduction pipeline with twilight flat division
and bias subtraction routines written in Python and using astrometry.net (Lang
et al. 2010) for image alignment and {\sc swarp} (Bertin 2010) for image 
co-addition.  Aperture photometry was calculated using {\sc sextractor} (Bertin \&
Arnouts 1996).
%{\bf , and the magnitudes are in the AB system.}

\subsubsection{NOT and LT}
The 2.5 m Nordic Optical Telescope (NOT) and the 2 m Liverpool Telescope (LT), located at Roque de los Muchachos
Observatory, La Palma in the Canary Islands (Spain), obtained three epochs of
photometry presented in this paper: 2013 September  ($r$ and $i$),
2013 September 3 ($i$), 2014 January 5 ($i$). Several
additional epochs of $griz$ NOT and LT photometry were obtained as part of a campaign
to observe SN~2013fu, which are presented in Cano et al. (2014). Image
reduction of the $r$ and $i$ photometric data was performed using standard
techniques in {\sc iraf}\footnote{{\sc iraf} is
  distributed by the National Optical Astronomy Observatory, which is operated
  by the Association of Universities for Research in Astronomy, Inc., under
 cooperative agreement with the National Science Foundation.}:  bias combine,
 bias-subtract, flat-field co-add, flat-field normalise, flat-field divide, align and co-add (Tody 1986,
1993).
  The optical data were calibrated using Sloan Digital Sky Survey (SDSS; Ahn
et al. 2012) stars in the GRB field, and converted into Vega magnitudes.
%Magnitudes are in the AB system.
%The magnitudes of NOT data in the Table attached to the electronic version of the paper are in the Vega system.  

%We obtained several epochs of $griz$ photometry with the 2.5-m Nordic Optical Telescope, the first occurring on 31-August-2013. {\bf Photometry obtained $\simeq143$ ks after the trigger in $r'$ and $i'$ bands is presented here, while the remainder is presented in Cano et al. (2014)}. The $ri$ optical data were calibrated using SDSS stars via aperture photometry in the GRB field, where a zero point between the instrumental and catalog magnitudes was calculated and applied in each filter.

\subsubsection{Skynet}
Skynet (Reichart et al. 2005) obtained images of the field of GRB~130831A on
2013 August 31 with four 17'' telescopes of the 
Panchromatic Robotic Optical Monitoring and Polarimetry Telescope
 (PROMPT) array at Siding Spring Observatory, New South Wales,
Australia. Further observations were taken on September 1, this time observing
also with the four 16" telescopes of the PROMPT array at the Cerro Tololo
Inter-American Observatory, Chile and the 41'' telescope at Yerkes Observatory, Wisconsin, USA. Beginning at 13:39 UT (T-T$_0 =  35.5$ min), exposures ranging from 60 to 180~s were obtained in the $BVRI$ (PROMPT), and $g'r'i'$ (Yerkes) bands. Bias subtraction and flat-fielding were performed by Skynet's automated pipeline. Post-processing occured in Skynet's guided analysis pipeline, using both custom algorithms and ones based on {\sc iraf}. Differential aperture photometry was performed on single and stacked images, with effective exposure times from 60 s to 2.7 h.
Photometry was calibrated against the catalogued $BVg'r'i'$ magnitudes of six
APASS\footnote{http://www.aavso.org/apass} DR7 stars in the field (Henden \&
Munari 2014). The calibrations for $RI$ magnitudes were derived using
transformations obtained from prior observations of Landoldt stars (A. Henden,
private communication). 
%The Skynet $BVRI$ magnitudes are in the Vega system, 
%while the $g'i'$ magnitudes are AB.

\subsubsection{IKI Network for Transients}
The field of GRB~130831A was observed by several facilities of the follow-up network organized by the Space Research Institute (IKI) of Moscow, where observations of GRBs are planned and data reduction is carried out. We detail the observations of GRB~130831A of the individual observatories below.

The International Scientific Optical-Observation Network (ISON; Molotov et
al. 2008, Pozanenko et al. 2013c) started observing on 2013} August 31 at 13:14:32 UT, i.e. $\sim 10$ minutes after the trigger, with the 0.65-m telescope SANTEL-650 \citep{volnova15185} of ISON-Ussuriysk observatory. Thirty unfiltered images, each with 120 seconds exposure, were taken.
The 50-cm telescope VT-50 of ISON-Ussuriysk observatory started to observe at 13:26:10 UT, 22 min after the trigger, taking 384 unfiltered images with exposures of 30 seconds in two epochs with a gap of about 1.8 h \citep{volnova15185}.
Starting at 19:12:30 UT during the first day, the 40-cm SANkovich-TELescope (SANTEL) -400AN of ISON-Kislovodsk observatory took 34 frames with exposures between 60 and 120 s \citep{volnova15188}.

The Astronomicheskii Zerkalnyi Telescope - 8 (AZT-8) 0.7-m telescope of Gissar (Tajikistan) observatory took 57 frames in $R$ band with 60 s exposure time each starting at 17:47:54 UT. 

The AZT-8 0.7-m telescope of the Chuguev Observational Station (Institute of Astronomy, Kharkiv National University) observed the afterglow in $R$ filter starting on September 1 at 19:37:49 UT (Volnova et al. 2013c).

%(14 frames of 120 s exposure, 14 frames of 100 s exposure, and 6 frames of 60 s exposure) 
The optical afterglow was also observed by the 1.5-m telescope AZT-22 of
Maidanak observatory (Uzbekistan) on 2013 September 1, 2, 4, 5, 7--11, 15, 16,
22, 27, and 29. Every observational night, frames were taken in the $R$
filter with an exposure time of 600 s each. On September 1 we also took
several frames in the $B$, $V$ and $I$ bands with the same exposure, and on
September 2 we obtained additional observations in the $B$ filter.  

Most of the  data obtained by Maidanak are presented in Cano et al. (2014) and used to study
SN2013fu; in this article we use the early data which were not contaminated by the
SN emission (see Sect. 3.3).

Observations were also taken with the 1.6-m telescope AZT-33IK of Sayan
observatory (Mondy, Russia) on September 3, 4, and October 10, 11 and 14,
taking several frames in the $R$-band with an exposure of  60 seconds each. The
optical afterglow was also imaged with the 2.6-m Shajn telescope of the Crimean
Astrophysical Observatory on September 5, taking several frames in the $R$ band
with an exposure of 120 seconds each. 

%and November 4--5, 

%These late data cover the SN epoch and are published in Cano et al. (2014).
All data obtained by the facilities indicated above were processed with the
same initial reduction including dark frame subtraction and flat-fielding, and
using the {\sc iraf} packages {\sc apphot} and {\sc daophot}.  For the
photometric calibrations we used four stars from the SDSS,
indicated in Table \ref{tab:referencestars}.
The $ugriz$ magnitudes of the reference stars were 
transformed to the $BVRI$ photometric system using the transformation equations attributed to Robert
Lupton in the SDSS online  documentation\footnote{www.sdss.org/DR7/algorithms/sdssUBVRITransform.html/\\Lupton2005}.

%All data produced by the ISON-network telescopes were processed using the IRAF packages APPHOT and DAOPHOT with a flat-field procedure. For the accurate photometric calibrations we used four SDSS stars selected using the automatic tool for secondary photometric standards selection. The $urgiz$ manitudes were transformed to the $BVRI$ photometric system using the Lupton transformation equations
%\footnote{IRAF  is distributed by  the National Optical Astronomy Observatory, which is operated  by the Association of Universities for Research  in Astronomy  (AURA)  under cooperative  agreement with  the National Science Foundation.} 

%{\bf The optical afterglow detected in the data obtained by 65-cm telescope SANTEL-650 
%of ISON-Ussuriysk observatory \citep{volnova15185}, was close to overexposure in the first 6 frames. To recover a flux from the OA on the first 6 frames we inscribed the PSF in the wings of the source profile excluding the central part. The PSF model was constructed using several non-overexposed nearby stars.}

\subsubsection{GTC}
The 10.4m Gran Telescopio Canarias (GTC; (Canary Island, Spain), equipped
with the Optical System for Imaging and Low-Intermediate Resolution Imaging
Spectroscopy (OSIRIS) instrument (Cepa et al. 2000),  performed deep $r^\prime$-band
imaging of the GRB field 13330~ks after the trigger (2014 February 1), in
order to obtain photometry for the host galaxy. Nine 60~s  images were
acquired in $2\times2$ binning, providing a pixel  scale of  0.25 arcsec
pixel$^{-1}$. 
%; {\bf thus the derived magnitudes are in AB system.} 
The images were dark-subtracted and flat-fielded using custom {\sc iraf}
routines. Aperture photometry was performed using {\sc daophot} 
tasks as implemented
in {\sc iraf}. Photometric calibration was based on SDSS standard stars present
in the OSIRIS un-vignetted field of view (7.8 arcmin $\times$ 7.8 arcmin).

\subsubsection{Results from other facilities}\label{radio}

In this subsection, we describe results obtained by other teams, not involved in our work. Nonetheless, we will use their results in the next sections.\\

{\bf Konus-Wind Observations.} GRB 130831A was observed by {\it Konus-Wind} onboard the {\it WIND} spacecraft.  Golenetskii et al. (2013) found that this burst had a duration of 35 seconds; it was detected up to $\sim6$ MeV. In addition, the spectrum is relatively soft; it can be fitted between 20~keV and 15~MeV with the Band model (Band et al. 1993), yielding a low-energy spectral index $\beta_1  = -0.61 \pm 0.06$, a high-energy spectral index $\beta_2 = -2.3 \pm 0.3$ and peak energy $E_{p} = 55 \pm 4 \mathrm{ keV}$). The fluence between 20~keV-10~MeV is $(7.6\pm0.4) \times10^{-6} $~erg cm$^{-2}$ (90\% CL).
We use the results of the {\it Konus-Wind} data analysis presented in Golenetskii et al. (2013), which spans a much wider spectral range than the BAT data, to derive the energetics of this burst. Thus, they are better suited to assess the energy emitted by the burst without using extrapolation. Using a cosmological k-correction (Bloom et al. 2001) and the redshift $z=0.479$ of this burst (Cucchiara et al. 2013), we derive 1--10000 keV rest-frame energetics of $E_\gamma = 1.06\times10^{52} $ erg. This burst follows the Amati relation (Amati et al. 2006; Amati et al. 2009; see also Cano et al. 2014, their Fig. 12). 

{\bf Radio.} GRB 130831A field was also observed by the Jansky Very Large Array (EVLA) and Combined Array for Research in Millimeter Astronomy (CARMA), Zauderer et al. 2013. According to the analysis reported by Laskar et al. (2013), no significant radio emission was detected: the $3\sigma$ upper limits are 38 $\mu$Jy and 71 $\mu$Jy at 5.8 and 21.8 GHz, respectively, at 0.64 days after the trigger. Similarly, CARMA obtained a 3$\sigma$ upper limit of 780 $\mu$Jy 0.76 days after the trigger (Zauderer et al. 2013).

\subsubsection{Building the UVOIR light-curves}

In the subsections above we summarized the observations of the optical
instruments and how the data produced by each one were reduced. 
%{\bf We present the data we have reduced and analyzed in a table that will be
%  linked to the online version of the article; a sample version is
%  Table \ref{UVOIR} }. 
We shall now describe how we combined these different datasets into a
homogeneous set of flux light-curves. Magnitudes were translated to fluxes 
using the zero-magnitude flux densities listed in 
Table~\ref{tab:fluxconversions}.
%{\bf converted into flux using the Gemini Observatory online magnitude to flux
%conversion tool\footnote{http://www.gemini.edu/?q=node/11119}}.
Our overall approach has been that when 
data have been obtained from multiple observatories in the same (or almost the
same) band, we scale all the data to match the
dataset which provided the largest number of measurements in that band. 
The largest set of measurements in $r'$ and $i'$ come from RATIR. We
normalized the NOT and LT $r'$ and $i'$ data points to RATIR using the
calibration stars common to the two datasets 
(see Table~\ref{tab:referencestars}). 
The comparison of the calibration
stars led us to scale the flux from the NOT $r'$ band data point by a factor
of 0.99 and, based on the scatter between the NOT and RATIR measurements of the
calibration stars we added a systematic error of 2 per cent to the errors on
the NOT fluxes. The NOT and LT $i'$ band flux was scaled by a factor of 1.01 and a
systematic error of 1 per cent was added.
The Skynet Yerkes $r'$ and $i'$ data were matched to RATIR using the common
reference stars listed in Table~\ref{tab:referencestars}. 
The $r'$ and $i'$ fluxes from Skynet were
scaled by factors of 0.94 and 0.91 respectively and systematics of 3 per cent
($r'$) and 2 per cent ($i'$) were added to the uncertainties on the Skynet 
fluxes.

The largest set of measurements in $B$, $V$, $R$ and $I$ come from the Skynet
PROMPT observations. To incorporate the UVOT $v$ and $b$ data into the $V$ and
$B$ light-curves, we first transformed the UVOT magnitudes of the GRB afterglow and the
Skynet calibration stars into Johnson magnitudes using the appropriate colour
transformations in Poole et al. (2008), before translating to flux densities. 
The UVOT and Skynet photometry of the
Skynet calibration stars 
(see Table~\ref{tab:referencestars}) were then compared to determine the
appropriate scaling factors and systematic errors for UVOT. The UVOT V and B
fluxes were scaled by factors of 1.13 and 1.07 respectively, and systematic
errors of 9 per cent and 6 per cent were added respectively to their flux
uncertainties.

Based on the photometry of the calibration stars in common between the
Maidanak and Skynet PROMPT observations 
(see Table~\ref{tab:referencestars}) in the $B$, $V$
and $I$ bands, we scaled the Maidanak $B$, $V$ and $I$ fluxes by factors of
1.01, 1.01 and 1.06 respectively and added systematic errors of 10, 5 and 7 per cent.
%No scaling was required for the Maidanak $V$ data, but a systematic error of 10 per cent was added to the Maidanak $V$ fluxes. 
For the $R$-band data from the IKI Gissar, Maidanak, Chuguev and Sayan
observatories, the calibration stars in common with Skynet suggested a
scaling factor of 1.01 for these data and a 6 per cent systematic
error, but the resulting GRB $R$-band light-curve from these
observatories appeared to be systematically lower than that obtained
from Skynet.  A power-law fit to the $R$-band flux light-curve derived
from IKI observations between 15 and 100 ks, performed simultaneously
with a power law fit to the Skynet $R$ band data in the same time
interval, and with the power-law slope tied between the two datasets,
gives a best-fitting normalisation for the IKI data which is 0.79
times that of the Skynet data. Therefore the IKI $R$ fluxes were
scaled by this factor before combining them with the Skynet PROMPT
data.

\begin{figure*}\label{2}

\begin{center}
%\epsscale{.80}
%\plottwo{GRB130831_optical.ps}{fluxcurve_final_new.ps}
\includegraphics[angle=-90,scale=0.70]{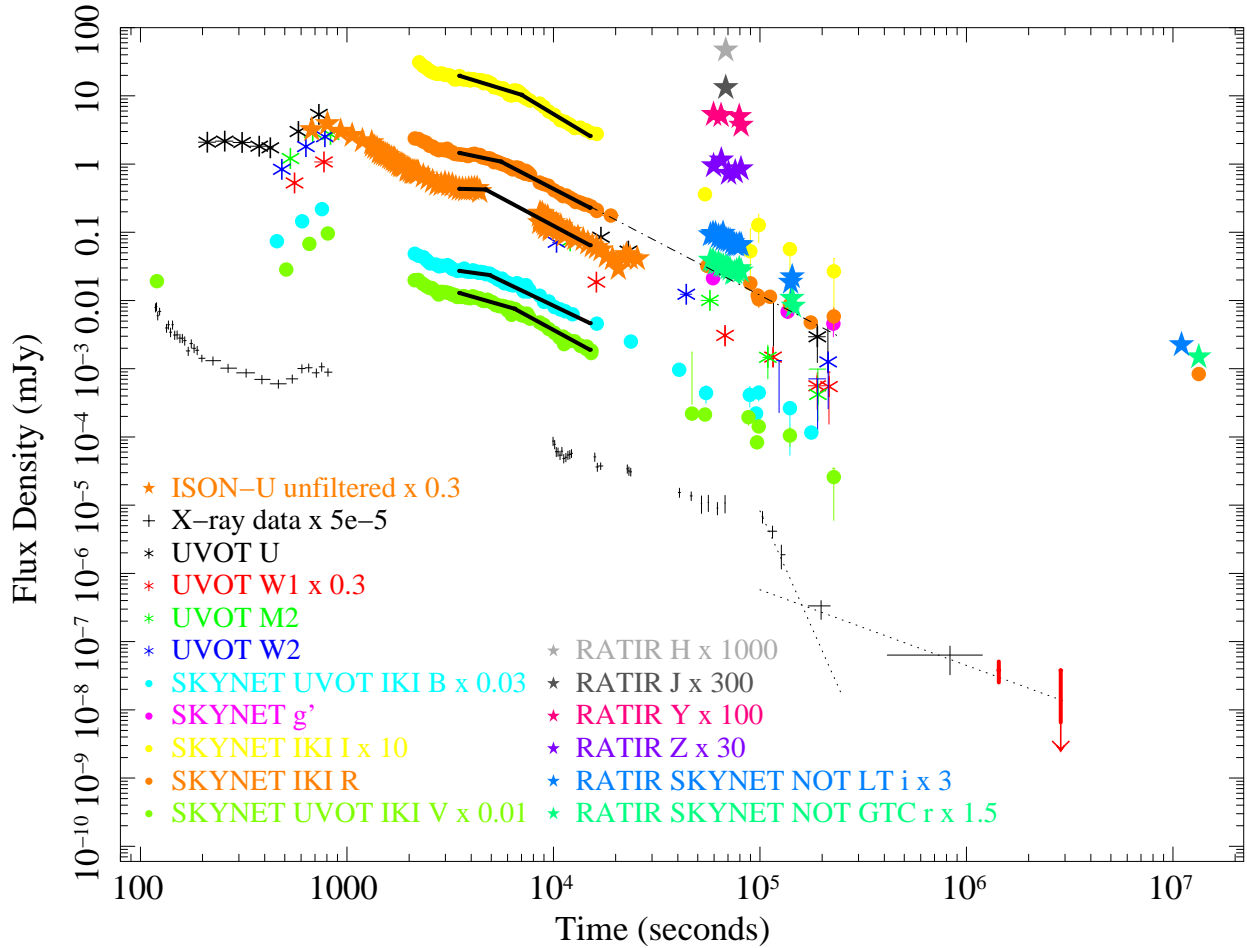}

\caption{GRB 130831A light-curves in near infrared, optical, UV and X-ray band. Black crosses are XRT data points, while red ones are {\it Chandra} data points. Data between 230 ks and 6000 ks are contaminated or dominated by the emission of the supernova associated with this GRB and are not shown here. Data points at $\simeq 10^7$~s show the brightness of the host galaxy. The reader is referred to Cano et al. (2014) for a complete study of the supernova. 
We plot the broken power-law models on the unfiltered and $BVRI$ light-curves between 3.5 and 15 ks (solid lines), the power-law model on the $R$ light-curve between 15 and 230 ks (dot-dashed line), the two power-law model on the X-ray that fits the data in this band after 100 ks (dotted lines). See table 2 and 6 for the values of the parameters. UV/optical/NIR data have not been corrected for Galactic and host galaxy extinction.}

\end{center}
\end{figure*}

\section {Results}\label{LCSEDS}

\subsection{Spectral analysis of X-ray data}\label{spectral analysis}

The XRT spectral data (0.3--10~keV) were extracted in nine different temporal segments, with boundaries selected according to the different observing modes and breaks in the light-curve (see Table 2). Ancillary response files and exposure maps were created using the {\sc heasoft} software for each segment, as well as appropriate response matrices from calibration database. All the spectra were fitted using {\sc xspec} v12.8.1 (Arnaud et al. 1996) with an absorbed power-law model. Some indication of spectral evolution was found, with the spectral index trending from soft, during the first orbit, to hard from the second orbit onwards. Detailed results are shown in Table \ref{tab:xrtspec}.

The X-ray spectrum between 9 ks and 132~ks, i.e. the end of the unusual late steep decay, can be modeled with a power-law with Galactic absorption and intrinsic absorption at the redshift of the burst, $z=0.479$. The Galactic absorption has been fixed to $N_{\rm H} = 4.8\times10^{20}$ cm$^{-2}$ (Kalberla et al. 2005). The best fit parameters are: spectral index $\beta_{\rm X} = 0.77\pm 0.07$, intrinsic $N_{\rm H} {\rm (z=0.479)} = 6.8 ^{+3.3} _{-3.1}\times10^{20}$ cm$^{-2}$ which is consistent with 0 at $2.1\sigma$ CL. 
 
 % = 6.8 ^{+3.3} _{-3.1}\times10^{20}$ cm$^{-2}$, which is consistent with 0 at $\simeq 2\sigma$ CL. 

%We requested to perform DDT observations with {\it Chandra} in form of 2 ToOs, which were granted. Following the approval of the request, {\it Swift} kept observing the GRB, and detected a dim X-ray afterglow. Such detection, however faint, indicated that the steep decay had broken into a slower slope. The two epochs {\it Chandra} DDT observations of GRB 130831A were on 2013 September 17 (T0+16.6 days) and 2013 October 3 (T0+33.1 days) for an exposure of 15 ks each.  The ACIS data were processed with CIAO v4.6 and CALDB v4.5.9, using standard tools.  We filtered the events in the ACIS-S3 chip for grades 0,2,3,4,6 and energy from 0.5-8 keV.  We extracted the source counts within a 1.5 arcsec radius region centered on the best known position of the GRB {\bf cite that position}, yielding 8 and 1 counts, respectively.  The first epoch yielded a detection with a significance of $5.4\sigma$ determined using the method described in Kraft, Burrows, and Nousek (1991).  The second epoch was only $\simeq 1\sigma$ above background. 

To infer the late X-ray flux from {\it Chandra} measurements, we applied the following procedure. First, we corrected for the portion of the PSF excluded from the 1.5 arcsec radius aperture. Using the PC mode XRT spectral fit parameters (spectral index $\beta=0.77$, intrinsic $N_{\rm H}=6.8\times 10^{20}$ cm$^{-2}$), we derived 0.3--10 keV fluxes for the two epochs of $7.4 \pm 2.5 \times 10^{-15}$ erg cm$^{-2}$ s$^{-1}$ and $7.8^{+18.0} _{-6.5} \times 10^{-16}$ erg cm$^{-2}$ s$^{-1}$, respectively. To derive the latter, we used the estimates on confidence limits for small numbers of events in astrophysical data (Gehrels 1986). The $3\sigma$ upper limit corresponding to the second {\it Chandra} observation is $7.4 \times 10^{-15}$ erg cm$^{-2}$ s$^{-1}$. Both XRT and {\it Chandra} fluxes are corrected for absorption. The results of the spectral analysis were used to compute the rate to flux conversion factors employed to build the flux light-curve.

\subsection{The X-ray light-curve of GRB~130831A.}\label{xraylc}

The X-ray light-curve is shown in Fig. 2. After an initial fast decay with slope $\alpha \simeq 6$ ending at $T_0+200$~s, there is an X-ray flare starting at about 500 s after the trigger and lasting at least up to the end of the first orbit at 900 s. The highest flux recorded was about $10^{-9}$ erg cm$^{-2}$ s$^{-1}$ at the beginning of XRT observations. When GRB observations resumed, $\simeq9$~ks after the trigger, the decay slope was slower than that at the very beginning of observations. This decay terminated at about $\simeq100$ ks after the trigger, when the flux showed a surprisingly steep drop.

 We fit the X-ray data (see Table \ref{tab: xrt time analysis}) from the very beginning with the sum of an early power-law, a broken power-law decay, a Gaussian flare superimposed on the second segment, and another, final power-law. The first power-law basically represents the tail of the prompt emission, probably due the curvature effect,
which then gives way to a shallower decay usually attributed to a different emission mechanism (Zhang et al. 2006). When this second process ends as well, the flux falls quickly again. The late power-law may represent emission powered by the FS mechanism, which can emerge once the emission from the previous process is over. The fit is acceptable, yielding $\chi^2/dof= 50.7 / 48$. We find a decay slope $\alpha_{X,2} = 0.8$ between $0.3$ and $98$~ks. Such a slope is intermediate between the typical shallow decay phase $\sim 0.3$ and the ``normal" decay phase $\sim 1.2$ seen in a wide sample of GRB afterglows (see Evans et al. 2009). We can still find a value of $0.8$, however, in the distributions of decay indices of both phases. After the break at $98$~ks, the best fitting temporal slope is $\alpha_{X,3} = 5.9^{+1.0} _{-0.4}$, much faster than is usually observed in late X-ray afterglows, even in the case of a jet break, when $\alpha \sim 2 - 3$ (Sari et al. 1999; Racusin et al. 2009). This steep drop suggests that the X-ray afterglow might have been, until then, produced by some internal dissipation mechanism rather than the typical FS emission. With this fit model, the late power-law component has a decay slope $\alpha_{X,4} = 0.90^{+0.11} _{-0.05}$. We note that this decay slope of the X-ray flux after $\sim200$ ks is steeper than the decay slope during the shallow decay phase. If the late X-ray flux were FS emission, it might have begun hundreds or even thousands of seconds after the trigger. However, the fitted model above does not include this possibility, and it may yield late power-law component slopes flatter than the real ones in order not to over-predict the flux at very early epochs. To better investigate the late emission, we use a different time interval. If we fit the X-ray data points from 100 ks onwards with a simple power-law model, we obtain a poor fit, $\chi^2/dof = 17.8/5$. The best fit decay slope is $\alpha_{X,3} = 4.6^{+0.5} _{-0.4}$. However, if we fit the same data points with a power-law + power-law model, we obtain a much better fit, with $\chi^2/dof = 2.4/3$. This model yields a decay slope for the first power-law of $\alpha_{\rm X,3} = 6.8 ^{+2.0}_{-1.5}$, with a $3\sigma$ lower limit (with $\Delta \chi^2 = 9$) of $\alpha_{\rm X,3} = 3.9$. Such a value, though, is still too steep for the FS model, see above. The best fit value for the late power-law component is $\alpha_{\rm X,4} = 1.11 ^{+0.22} _{-0.29}$, while its flux at 2 days (173 ks) after the trigger is $6.08 ^{+2.31} _{-2.77} \times10^{-14}$~ erg cm$^{-2}$~s$^{-1}$.
  
An analysis with the F-test suggests that the two-power-law model is not necessary, because the probability of an improvement by chance is $4\%$, which is not negligible. However, if we adopted the simple power-law model with the steep decay slope $\alpha = 4.6$ after 100 ks, then the flux at the time of the first {\it Chandra} observation would be $\simeq 5\times10^{-17}$ erg cm$^{-2}$~s$^{-1}$. Such extremely low flux corresponds to less than $0.1 $ counts with a ACIS-S 15 ks observation; thus our first {\it Chandra} observation would most likely yield 0 counts. We can place a 99.5\% CL upper limit of 5.3 counts, according to Gehrels (1986). This prediction, however, is in disagreement with the fact that the observation actually produced 8 counts. According to Kraft et al. (1991), these 8 counts represent a $5.4\sigma$ detection. We therefore conclude that the X-ray afterglow decay has become much shallower at late epochs, and we will adopt the results of the fit of the two-power-law model after 100 ks.

\subsection{Optical light-curve of GRB 130831A}
The combined optical light-curves in Fig. 2 show an initial short plateau, which lasts until $\sim 500$ s, followed by a steep rise and a peak at $\simeq 800$~s. This optical flare is basically concurrent with the X-ray flare. Following the flare, there is another plateau that in turn gives way to a steeper decay at $\simeq 5000$~s. In this phase, fitting the optical light-curves with simple models such as a broken power-law does not provide a statistically acceptable fit. For example, the best fit of $B$ band data between 3.5 and 15 ks with a broken power-law yields an early decay slope of $\alpha_1 = 0.42 \pm 0.08 $, break time $t_\mathrm{break} = 4.90 \pm 0.08$~ks, and post-break slope $\alpha_2 = 1.45 \pm 0.03$ with $\chi^2 /dof = 81/49$. Fitting the other light-curves in this interval yields similar results. We nonetheless plot the best-fit curves in Fig. 2, as an indication for the behaviour of the optical afterglow, and report  the results of fitting the light-curves in the interval between 3.5 and 15 ks in Table \ref{tab: opt time analysis}.
Such high $\chi^2$ are due to some ``wiggles'' of the densely sampled light-curves in this phase, which has been seen in other GRBs (e.g. Swenson et al. 2013; Matheson et al. 2003). The post-plateau decay is moderately steep and does not seem to change its slope at the epoch of the X-ray drop. However, about $\sim5$ days after the trigger, the optical emission starts to rise due to light coming from SN 2013fu, the supernova associated with GRB 130831A (Klose et al. 2013; Cano et al. 2014). Given the complication of considering the SN flux, we have excluded all data points that had a $> 10\%$ contribution from the SN, basically those after $\sim 230$~ks and before $\sim 6000$~ks. The SN contribution at these epochs have been estimated using the SN 1998bw template program presented in Cano (2013). \\ In addition, we have observations at $T-T_0>100$ days in $r'$ and $i'$ filters, taken with Gran Telescopio Canarias (GTC) and LT (see Sect. 2.2.3 and 2.2.6). These late time data do not suffer significant contamination from the SN and the afterglow, which have faded away. They correspond to the magnitude of the host galaxy, and have been used to determine the optical afterglow behaviour (see below). Vega magnitudes of the host galaxy are $r' = 23.75\pm 0.11$, $i' = 23.83 \pm 0.10$. Assuming the conversions from Jordi et al. 2006 (their Table 1), we find a Vega magnitude $R = 23.84 \pm 0.15$. Spectral observation taken at the Gemini North Observatory revealed that GRB 130831A occurred at redshift of $z=0.479$ (Cucchiara et al. 2013). For such a redshift, the magnitude of the host corresponds to a luminosity $L\simeq 0.04 L_{\star}$ in the B band (Hjorth et al. 2012).

% assume that these observations basically provide the luminosity of the host galaxy. Its magnitude, in the R band, is $R=23.65 \pm 0.15$.

We know the host contribution in $R$, $r'$ and $i'$ only, from our late-time GTC and LT images. We fitted the light-curves in these filters with the same model, namely a power-law $F \propto t^{-\alpha}$ plus constant, from 15 ks up to 100 days after the trigger. We ignored the data before 15 ks because they would lead to a very bad fit, as we previously noted (see above). The best fit decay slopes and $\chi^2$ are $\alpha_R = 1.60\pm0.03$, $\chi^2 /dof = 9.4/10$; $\alpha_{r'}=1.49 \pm 0.06$, $\chi^2 / dof = 30.7/14$; and $\alpha_{i'} = 1.64 \pm 0.07$, $\chi^2/dof = 29.2/15$ (see Table \ref{tab: opt time analysis}). These fits are statistically acceptable and consistent within $2\sigma$. We find that the weighted mean is $\alpha_{\rm opt} = 1.59 \pm 0.03$. The decay slopes of the flux in other filters are consistent with this value within $3\sigma$ as well. We note that Cano et al. (2014), in their analysis of the afterglow light-curves, find that the optical decay slope is $\alpha=1.63\pm0.02$, consistent with our analysis. We remark upon the fact that we can fit the optical light-curves with an uninterrupted power-law, even across the X-ray break. This feature strongly suggests that optical and X-ray emission (at least part of it) have different origins.

\subsection{Spectral Energy Distributions}

To test the hypothesis that the late emission is entirely due to FS, we built a spectral energy distribution (SED) with the available UVOIR + X-ray data at 2 days after the trigger (173 ks).

First, we calculated the count rates at 2 days. We used the data between 15 ks and $100$ ks, since no colour evolution was detected in the UV to the near IR, and the count rate in all light-curves could be fitted as a power-law with a common decay index $\alpha=1.59$ between these two epochs. The UVOT data were translated to {\sc Xspec}-compatible files using the standard {\sc ftool} {\sc uvot2pha}. Then, we adjusted the count rates of these files to the values determined by fitting the light-curves. Each of the ground based optical and near IR photometric data points were imported into {\sc Xspec} using bespoke software, as follows. Each photometric data point was recorded as a single-channel spectral file containing a count rate and count-rate uncertainty, with a corresponding response file. To produce the response files, the responsivity of the filter/telescope combination as a function of wavelength was converted to a normalised effective area as a function of energy. As for the X-ray, we first determined the light-curve count rate at 2 days, $fit_{CR}$. We then determined a new exposure time $t_{newexp}$ for which $spec_{CR} / t_{newexp} = fit_{CR}$, where $spec_{CR}$ is the count rate of the source after background subtraction. We then imported the source and background XRT spectral files with the changed exposure times into {\sc Xspec}. To build the SED, we used the XRT data after the steep decay slope, but did not use the {\it Chandra} data because XRT and {\it Chandra} fluxes would have to be renormalized to the same value and XRT has more counts.

\begin{figure*}
\begin{center}
\includegraphics[angle=-90,scale=0.60]{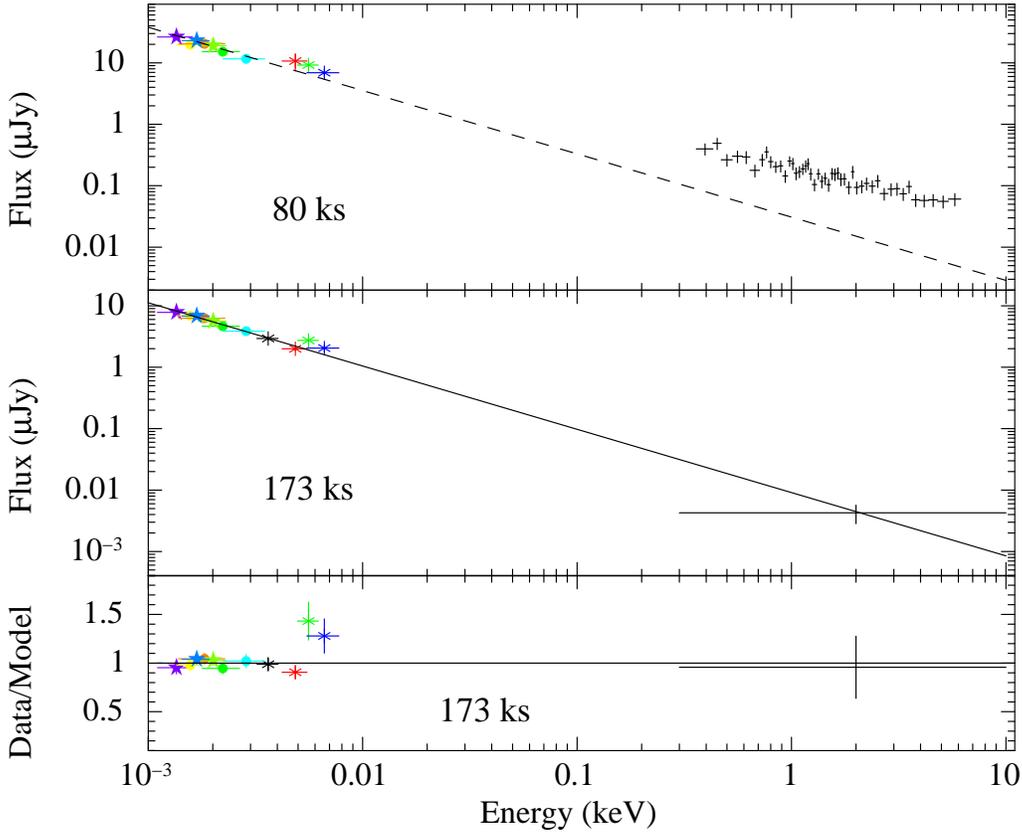}
\caption{Spectral Energy Distributions of GRB 130831A at 80~ks (top) and 173 ks (2 days) after the trigger (bottom).
At 173 ks, we plot on the SED the best-fit model, a simple power-law model with $\beta_{OX}=1.03$. We rescaled  this best-fit model, multiplying the normalization constant by $(173/80)^{1.59}$, where 1.59 is the temporal decay slope, and we plot such an ``extrapolated model" with a dashed line on the 80 ks SED. Such an extrapolation predicts the optical but clearly under-estimates the X-ray emission, which must be produced by a component not present at 173 ks. Each filter has the same colour in the plot.}
\label{SED}
\end{center}
\end{figure*}

 % We then found the count rate for the source only, using the formula $spec_{CR} = C_{tot}  - C_{bkg} \times A1/A2$, where $C_{tot}$ is the total (source+background) count rate, $C_{bkg}$ is the background count rate, A1 and A2 are the sizes of the of the aperture used when extracting source and background. To get the values of $C_{tot}$, $C_{bkg}$, A1 and A2 we used the {\rm ftools} commands {\rm fstatistic} and {\rm fkeyprint} on the source and background spectral files. Knowing the values of $spec_{CR}$, we could determine a new exposure time $t_{newexp}$ for which $spec_{CR} / t_{newexp} = fit_{CR}$. We then changed the exposure times of the source and background spectral files to $t_{newexp}$, using the {\rm fparkey} command. 
%We do not use the data after $\sim230$~ks to avoid contamination from SN 2013fu.%To subtract the residual contamination from the SN, we obtained the expected flux from it following Cano et al. (2013) in the filters XX XX XX and subtracted it from the data points.  For the remaining filters, we did not include data later than 100 ks and extrapolated them up to 173 ks.
We fitted the optical and X-ray data with {\sc Xspec} (Arnaud et al. 1996).  We adopted a simple power-law model with two absorbers and two {\it zdust} components, one at $z=0$  and another one at $z=0.479$, i.e. the redshift of the burst (Cucchiara et al. 2013). The values of the absorbers are the same given in the X-ray data analysis section \ref{spectral analysis}. The Galactic reddening was fixed at $E(B-V)=0.04$ mag according to the map of Schlegel et al. (1998). As for the extinction at the redshift of the burst, we tried the Milky Way, Large and Small Magelanic Clouds (MW, LMC and SMC) extinction laws as in Pei (1992). We found that all of them yield acceptable and similar fiting results. However, the MW extinction law provides the best fit, so we have adopted the results of the fit with this law. 
The SED of GRB 130831A at 2 days is shown in Fig. \ref{SED}. The best fit parameters are an index $\beta_{\rm OX}=1.03^{+0.05} _{-0.04}$ and a small or absent amount of extragalactic reddening $E(B-V) = 0.02\pm0.01$ mag (Cano et al. 2014 find a negligible rest-frame extinction as well); this fit yields $\chi^2/\rm{dof} = 12.7 / 9$. A broken power-law model does not significantly improve the fit, since it yields $\chi^2 = 12.4 / 8$ and the break energy is unconstrained. Moreover, fits with LMC and SMC extinction law result in a break energy above the X-ray band. All in all, we believe that a simple power-law model is adequate to describe the SED at this late epoch. Results of the fit of the 173 ks SED are shown in Table \ref{SEDfit}. There are only $\simeq 15$ counts collected by XRT after the fast drop, and to build the SED and obtain the quoted results we constructed a single X-ray data bin spanning from 0.3 to 10 keV. We were concerned that the use of such a wide bin might be not the optimum in the SED fitting process. Therefore, we repeated the fit using standard $\chi^2$ statistics with optical data and Cash statistics for the X-ray data, where the bins were constituted of single counts. We obtained very similar results. The fit with a simple power-law model yielded $\beta_{OX} = 1.03 ^{+0.05} _{-0.04}$, $E(B-V) = (1.9 \pm 1.3) \times 10^{-2} $ mag with total statistics of $23.4$ and 26 {\rm dof}. A fit with a broken power-law model was marginally better, yielding statistics of 22.2 with 25 {\rm dof}, but the F-test indicates that the probability of an improvement by chance was high, with a probability of $25\%$.

We also show that the same model does not apply to earlier data. Following the same procedure outlined above, we built a SED with UVOIR and X-ray data at 80 ks (Fig. \ref{SED}), before the steep X-ray drop. Then, we changed the normalization of the 173 ks power-law fit, following a decay slope of $\alpha=1.59$.  We then plotted such a re-normalized power-law model with $\beta_{\rm OX} = 1.03$ onto the 80 ks SED. We see that, while the optical emission is easily matched, the observed X-ray flux lies well above the model prediction. This finding is confirmed by the fact that if we fit the 80 ks SED with the same power-law model used for the 173~ks SED, we obtain a best-fit spectral slope $\beta_{\rm OX} = 0.76 \pm 0.01$. This is harder than the slope found at 2 days and inconsistent with it. This result confirms that the spectrum at 80 ks across the X-ray and optical bands is not consistent with the spectrum at 173~ks and lends credence to the idea that there is an additional component in the X-ray band at early epochs.

\section{Discussion}

\subsection{Modeling of GRB 130831A and the Efficiency of its ``Central Engine".}

We found that the X-ray light-curve, after the steep drop at 100 ks, resumes a slower decay slope of $\alpha_{\rm X,4} = 1.11^{+0.22} _{-0.29}$; such a decay slope is statistically consistent with the optical decay slope from 15 ks onwards, $\alpha_{\rm opt} = 1.59\pm 0.03$, at $2.1\sigma$ CL. The SED at 2 days (173 ks) after the trigger is adequately fit with a simple power-law model, in which the X-ray and the optical band lie on the same spectral segment. Values of the spectral and temporal indices seem to be typical of GRB late afterglow emission, and are easily explainable by the FS model.\\
The FS model predicts basic relations between spectral and decay indices (for a review, see Zhang et al. 2006), which depend on the type of expansion (spherical or jet) and the spectral regime, i.e. where the observing bands are located relative to the synchrotron peak frequency and cooling frequency, $\nu_{\rm m}$ and $\nu_{\rm c}$ respectively, and the density profile of the circumburst medium. We can use these closure relations to find the conditions that apply to the case at hand. We adopted $\alpha_{\rm opt}$ as decay slope, since it is better constrained than the late X-ray decay index, and $\beta_{\rm OX}$ as the spectral slope. If the observing bands (both the X-ray and optical) were below $\nu_{\rm c}$ and the medium had a constant density profile, we would have $\alpha -  \frac{3}{2} \beta = 0$; this is in agreement with observations at $1\sigma$. If the medium had a stellar wind profile, with density $\rho$ decreasing with radius $r$ as $\rho \propto r^{-2}$, then $\alpha -  \frac{3}{2} \beta - 1/2 = 0$; this is ruled out at $\simeq 10\sigma$ level. If the observing frequency is above the cooling frequency, for both the constant medium and stellar wind profile cases the relation $\alpha - \frac{3}{2}\beta +1/2 = 0$ should be satisfied; but this is rejected at $\simeq 10\sigma$. Finally, if the outflow is collimated and has decelerated enough so that the observer detects emission from the edges, the decay should become steeper (``jet break"). After the jet break, the relations to satisfy are $\alpha -2\beta -1 = 0$ and  $\alpha -2\beta = 0$ if the observing frequency is below or above $\nu_{\rm c}$. These relations are ruled out at $16\sigma$ and $5\sigma$ respectively. The only relation of those above fulfilled within $1\sigma$ is that for the observing frequency below $\nu_{\rm c}$, constant density medium, and pre-jet break expansion.

%that fits the observed spectral and temporal indexes is $\alpha = \frac{3}{2} \beta$, which is expected for $\nu_{\rm m}<\nu_{\rm obs}<\nu_{\rm c}$ and the shock moving in a medium with constant density.
%such condition is satisfied within $1\sigma$. 

In the context of the FS model, the flux produced by the afterglow depends on several factors, such as the kinetic energy $E_K$ of the outflow, the fraction of energy given to radiating electrons $\epsilon_{e}$ and to the magnetic field $\epsilon_{B}$, the density of the environment $n$, the slope of the power-law energy distribution of electrons $p$, the type of expansion and where the observing bands are located. It is possible to derive the kinetic energy through simple relations once the flux is known, but one has to make assumptions of the values of the other parameters.

Following Zhang et al. (2007), the flux density $F_{\nu}$ of the FS emission is given by 

\begin{eqnarray}\label{eq1}
F_{\nu} (\nu_m < \nu < \nu_c) = 1600 D_{28} ^{-2} (1+z)^{\frac{3+p}{4}} \epsilon_{B,-2} ^{\frac{p+1}{4}} \epsilon_{e,-1} ^{p-1}  \times \nonumber \\ 
E_{K,52} ^{\frac{3+p}{4}} n^{1/2} t_d ^{\frac{3}{4}(p-1)} f_p ^{p-1} \left( \frac{\nu_{3.3\times10^{12}}}{\nu} \right) ^{\frac{p-1}{2}}~{\rm \mu Jy}
\end{eqnarray}
Where $f_p$ is a parameter depending on $p$, and $D$ is the luminosity distance in cm. Subindices indicate normalized quantities, $Q_{x} = Q / 10^{x}$ in cgs units. A spectral index $\beta_{\rm OX} = 1.03$ for an observing frequency $\nu$ between $\nu_m$ and $\nu_c$ implies $p=3.06$. Such a large value of $p$ is not very common; typically $p = 2.1 - 2.5$. However, $p=3.06$ is still within two standard deviations from the centre of the distribution of this parameter, as implied from the analysis of more than $300$ {\it Swift} GRBs (Curran et al. 2010), and we will adopt this value. As for the observing frequency, we take $\nu=2.42 \times10^{17}$~Hz, i.e. 1 keV. We convert the X-ray flux between 0.3 and 10 keV into a flux density of $\simeq 7\times10^{-3}$~$\mu$Jy  at this frequency. 
% we take $\nu=3.3 \times10^{14}$~Hz, i.e. the $z'$ band, where we obtain a flux of $7\mu$Jy 2 days after the trigger. We choose this infrared band, rather than the X-ray band, in order to avoid the uncertainties in the conversion of the X-ray flux into flux density caused by the errors in the spectral index. Moreover, this burst is subject to negligible extinction and our choice of the $z'$ band further minimizes the effect of such extinction.

%We convert the X-ray flux between 0.3 and 10 keV into a flux density of $1.84\times10^{-3}$~$\mu$Jy at this frequency. 
%{\bf I wonder, couldn't we then use the optical flux, rather than the X-ray flux, to this aim? After all, extinction is low and we have some NIR data.}

From the equation above, we obtain

\begin{equation}\label{eq2}
\epsilon_{B,-2}^{1.02} \epsilon_{e,-1} ^{2.06} E_{K,52}^{1.52} n^{1/2} \simeq 2.0
\end{equation}

We infer $E_{K,52} = 11.8 $ if we assume $\epsilon_{e,-1} = 2.7$, $\epsilon_{B,-2} = 0.2$, and $n=10^{-3}$. These low values of $\epsilon_B$ and $n$ are required to have $\nu_{c}$ well above the X-ray band 2 days after the trigger. A lower value of $\epsilon_B$ may imply significant Inverse Compton flux, which is usually not detected in GRBs. The low density is not totally unprecedented in GRB afterglow modeling, since it has already been found for events in constant density media (Panaitescu \& Kumar 2002, Cenko et al. 2011; with radio observations) and stellar wind media (Cenko et al. 2011, Perley et al 2014). On the other hand, a density {\it lower} than $\sim 10^{-3}$ is not usually expected in long GRBs, which occur next to, or within, dense star forming regions. We also note that $E_K$ has a weak dependance on $n$, so our inferred $E_K$ would not be dramatically different if the value of this parameter were within one order of magnitude from what we use. As for $\epsilon_{e,-1}$, the value we have chosen is close to that of equipartion, and is obtained by modeling GRB afterglows (see aforementioned references).
However, we now show that in our modeling we cannot have $\epsilon_{e,-1} \ltsim 2.5$. From Eq. \ref{eq2}, we have $E_K \propto\epsilon_e^{-4/3}$ roughly. FS theory predicts that $\nu_m \propto E_K^{1/2} \epsilon_e^2$ and $F_{\nu} (\nu_m) \propto E_K$, where $F_{\nu} (\nu_m)$ is the peak synchrotron flux reached at $\nu_{\rm m}$. For $\nu<\nu_{\rm m}$, $F_{\nu} \propto \nu^{1/3}$; the flux below $\nu_{\rm m}$ will therefore be $F(\nu) \propto \epsilon_e^{-16/9}$. For the values of the parameters quoted above, the radio flux at 22 and 5.8 GHz predicted 0.67 days after the trigger would be 56 $\mu$Jy and 36 $\mu$Jy respectively, below the upper limits determined by the VLA (Laskar et al. 2013). However, if $\epsilon_{e,-1}$ were less than $2.5$ the predicted radio flux would instead be larger than the above limits. We note, though, that interstellar scintillation could suppress the observed radio flux as well. For the values of parameters we've constrained, the FS peak flux will be $\simeq550$~$\mu$Jy. Such a flux is quite typical for GRB FS peak; see Chandra et al. (2012) and De Ugarte-Postigo (2012). We can assume that our estimate on $E_K$ is robust at least to an order of magnitude.

The modeling above enables us to determine the efficiency $\eta$ of the conversion of the energy of the outflow into energy $E_{\gamma}$ emitted in high energy photons during the prompt emission. The efficiency is defined as $\eta = E_{\gamma} / (E_{K} + E_{\gamma} )$. Given $E_{\gamma}=1.06\times10^{52}$~erg and the value of $E_K$ constrained above, we find $\eta \simeq 0.07$. By means of both optical and X-ray data, we have thus measured the efficiency of the central engine of a GRB with a prolonged internal emission episode. The optical light-curves were well sampled, and late {\it Chandra} observations constrained the X-ray flux light-curve after the steep drop. Thus, we could establish with little doubt that the late emission was entirely consistent with the FS model. Zhang et al (2007) and, more recently, L\"u \& Zhang (2014) carried out a similar task by using XRT light-curves; such a study with X-ray data alone, especially without precise late measurements, may be more ambiguous.

An efficiency of $\eta \simeq 0.07$ is lower than that of those few GRBs that unambiguously show a plateau of ``internal origin" (see figure 12 of L\"u \& Zhang 2014). The efficiency of GRB 130831A is more characteristic of those GRBs that have their afterglow emission entirely explained by the FS mechanism, but with the presence of energy injection into the ejecta perhaps powered by a magnetar. Typically, GRBs with a plateau of internal origin have $\eta \sim 0.5-1$, while those explained by FS have a range of $\eta \sim 0.001-0.1$ with most clustering around $\eta = $ a few $\times$~$0.01$. 

%{\bf We note that the small density $n$ implies a relatively weak FS flux (see Eq. \ref{eq1}). This condition may explain why even a relatively weak non-FS emission (see next section) overwhelmed it.}

%An efficiency of $\eta \simeq 0.07$ is intermediate between those few GRBs that unambiguously show a plateau of ``internal origin" and that of GRBs that have their afterglow entirely explained by FS emission (see figure 12 of Lu \& Zhang 2014). Typically, the former have $\eta \sim 0.5-1$, while the latter have a range of $\eta \sim 0.001-0.1$ with most clustering around $\eta \simeq$ few $10^{-2}$. 

\subsection{Origins of the X-ray radiation between the end of the prompt emission and up to the 100 ks drop.}\label{origin}

\subsubsection{Observations}

The sudden drop in the X-ray flux at $\simeq 100$ ks, with a decay index $\alpha_{\rm X,3} \simeq 7$, cannot be interpreted as FS emission. The steepest decay in this model is $\alpha=p$, where $p$ is the index of the power-law energy distribution of radiating electrons, which occurs during a jet-break expansion phase. However, $p \simeq7 $ is not predicted at all on theoretical grounds (e.g. see Rieger et al. 2007); thus it is very difficult to explain such a steep decay index at late times. Theoretically, an index as steep as $\alpha \sim 3$ can be achieved by taking into account relativistic effects in simulations (e.g. Granot 2006, van Eerten et al. 2010) even if $p<3$, but the value of the decay index reached in the case of GRB 130831A is greater than this prediction. Duffell \&MacFadyen (2014) explored the possibility that the plateaux we see in GRB afterglows are produced by a jetted outflow before deceleration, followed by a steeper decay, which flags the Blandford \& McKee deceleration of the ejecta. However, such a decay in their model does not reach a value as steep as $\alpha\sim 7$ detected in GRB 130831A between 100 and 200~ks.

By assuming that the steep decay is due to the curvature effect (Kumar \& Panaitescu 2000 ; see also Uhm \& Zhang 2014), we followed Liang et al. (2006) to test the internal origin of X-ray emission up to the steep break. We found that to satisfy the $\alpha \simeq \beta+2$ condition of the curvature effect, the zero time $T$ of this emission needs to be $\simeq 75$ ks after the trigger, slightly before the beginning of the flux drop. The very steep decay component therefore strongly suggests that the X-ray emission up to 100 ks is of internal origin, since $T$ is allowed to occur at large time intervals from the initial prompt emission in this case (Zhang et al. 2006; Liang et al. 2006). 

In a more general study of the curvature effect emission (Uhm \& Zhang 2014), the emitting ejecta are assumed to accelerate or decelerate while producing the radiation. If the ejecta are accelerating, which may be the case for the magnetically dominated jet of the ICMART model (Zhang \& Yan 2011), the decay index may temporarily reach a value of $\alpha \sim 7$ we observe even assuming $T$ as the trigger of the prompt emission and $\beta \sim 1$. A magnetically dominated jet thus appears to be a reasonable solution for the GRB at hand. 

However, whichever solution applies, one would always have to assume that the emission is produced inside the ejecta and not in the medium surrounding the explosion as in the case of the FS scenario. Finally, in the previous section we made it clear that the final part of the X-ray light-curve, after the end of the steep decay, appears to be FS emission.

%Furthermore, the optical emission decays with a much more typical slope $\alpha=1.58$, explained by the FS emission, and shows no break at the time of the steep drop of the X-ray flux. Instead the {\it late} X-ray slope, as mentioned above, is statistically consistent with the optical slope. Finally, the SED built with the late X-ray and optical data shows that these two emissions are on the same spectral segment, and the spectral index $\beta_{\rm OX}=1.03$ is in agreement with the prediction $\alpha=\frac{3}{2}\beta$ of the FS model. \\
{\it We conclude that the X-ray emission up to the 100~ks break is not produced by the external, circumburst medium energized by the FS, but it is instead of ``internal origin", generated directly within the explosion outflow. This component stops abruptly and the X-ray flux drops rapidly, until the FS emission in the X-ray band prevails.} The optical emission is basically dominated by the FS mechanism (with the possible exception of the early flare; see Sect. \ref{early}). In the next sections, we shall discuss the possible origins of the high-energy emission between the end of the prompt emission and the fall of the X-ray flux at 100~ks.

\subsubsection{A Magnetar Central Engine}
The nature of the X-ray afterglow and the nature of the central engine are two of the many open questions in the contemporary field of GRBs (see Zhang 2011 for a review), even more so because this GRB shows that we may be misinterpreting the behaviour of other X-ray light-curves solely or primarily attributed to FS. For example, if we had not observed the steep drop at 100~ks for 130831A, we might have easily mistaken the relatively ordinary decay and spectral slopes as being produced by the standard FS-emission.

The core of the stellar progenitor of GRB 130831A may have collapsed into a magnetar (see for example Thompson et al. 2010). The rotational kinetic energy $E_{rot}$ of such objects is

\begin{equation}
E_{\rm rot} = 3\times10^{52} \left(\frac{M}{1.4\,M_{\odot}}\right) \left(\frac{R}{\rm  12\, km}\right)^2 P_{\rm ms} ^{-2}~\rm{ erg}
\end{equation}
where $M$, $R$ and $P$ are the mass, radius and period in ms of the object, respectively. Assuming unitary values for the parameters above, $E_{\rm rot}\simeq 3\times10^{52}$ erg, which is enough to power the supernova explosion and ultra-relativistic outflow.  If the mass M is closer to the Tolman - Oppenheimer - Volkoff limit, which is thought to be slightly larger than $2 M_{\odot}$ (Antoniadis et al. 2013), the parameter $E_{\rm rot}$ might be slightly different (Metzger et al. 2015).

In a simple scenario, the magnetar would initially tap into rotational energy and produce a very energetic outflow, likely roughly collimated into bi-polar jets. Such a wind would be produced through the process of dipole spin-down. The energy of the outflow is initially imparted to the stellar envelope, causing, or at least contributing to, the supernova explosion. Moreover, the magnetar outflow may be long-lived, and produce radiation that we observe (see below). In this scenario, the luminosity $L_0$ of the magnetar could be roughly constant, even for a relatively long timescale $T_{\rm em}$ (Zhang \& M\'esz\'aros 2002), depending on physical parameters. After $T_{\rm em}$, or if the magnetar collapses into a black hole, the light-curve would show a flux drop. This magnetar model is relevant for a few GRBs that show a plateau with approximately constant flux, with $\alpha \simeq 0$, followed by the steep slope segment (see cases studied by Liang et al. 2007, L\"u \& Zhang 2014, Cano et al. 2014, Bernardini et al. 2014, L\"u et al. 2015). However, the early decay slope of 130831A is $\alpha_X \simeq 0.8$, which is in contrast to the aforementioned cases, and requires a more complicated model to explain our observations.\\

\subsubsection {A Magnetar with decaying magnetic field}
In the simple spin-down calculation above, one assumes that the magnetic field $B$ is constant and independent of the period of the magnetar $P$. Metzger et al. (2011), however, assume that $B$ is linked to $P$, taking into account that the magnetic field is generated from the energy available in differential rotation. They thus estimate
%However, this may not be true. 
%It is reasonable to assume that magnetic field will decay with time; on the other hand, the period will increase with time since the magnetar is converting its rotational energy into the energy of the outflow. The luminosity $L$ of the spin-down process goes $L \propto B^2 P^{-4} \propto t^{-a}$; if $B$ depends on time as $B \propto t^{-q}$, then $P \propto t^{2q-a}$. In our case, $a=0.9$. Metzger et al. (2011) have found that, for $q=a$, one can can retrieve

\begin{equation}\label{magnetic}
%    B = 7\times10^{15} \epsilon_{B,-3} R_6 ^{-1/2} P_{-3}~\mathrm{G}
B = 10^{16} R_6 ^{-1/2} P_{-3} ^{-1} ~\mathrm{ G}
 \end{equation}

Under this assumption, $B$ decreases with increasing $P$, i.e., with increasing time since the explosion. In such conditions, the jet luminosity $L$ is not constant. As shown in Figures 2 and 5 of Metzger et al. (2011), $L$ decreases with time, scaling approximately as $L\propto t^{-1}$ from 100--1000 s to tens of ks after the collapse for reasonable values of parameters at the beginning of the spin-down, such as $B_0 \simeq10^{15}$ G and $P_0 = 1 - 2$~ms. The predicted luminosity of the jet seems to be in the right range to explain the X-ray emission of GRB 130831A during the slow decay phase, too. For the values quoted above, Metzger et al. (2011) predict a few $\times 10^{46}$ erg s$^{-1}$ at 10 ks; at the same epoch, GRB 130831A had a luminosity of $\sim 10^{46}$ erg s$^{-1}$.
%\footnote{ Inferred from the 0.3-10 keV flux of $\sim10^{-11}$ erg cm$^{-2}$ s$^{-1}$ at 10 ks, and a redshift z=0.48.} 

We note that a jet luminosity $L_{jet}$ does not convert immediately into X-ray luminosity $L_X$; one has to take into account that the radiation mechanism will have a certain efficiency. Moreover, we have not yet considered that GRB emission is beamed. We can write
% the true, beamed-corrected luminosity can be lower than that one gets assuming isotropic emission. 

\begin{equation}
 L_X = \eta_X f_{b,x}^{-1} L_{jet}
 \end{equation}

where $\eta_X$ and $f_{b,x}$ are the efficiency in converting the jet luminosity into X-ray radiation and the correction for the beaming, respectively.

% $L_{corr}$

According to Metzger et al. (2011), at late epochs ($>100-1000$~s after the collapse), the magnetar outflow is highly relativistic and Poynting-flux dominated; in such conditions, internal shocks and reconnections within the jet itself are not possible. Forced reconnection, however, can occur at large radii, when the outflow collides with the circumburst medium and/or the previous ejecta, and convert the jet energy into X-ray emission. Assuming that the efficiency of this process is similar to the one which generates the prompt emission, and the correction for beaming is 10 times lower than that during the prompt emission, Metzger et al. (2011) find that they can explain the observed $L_X$ and plateau durations of several GRBs similar to GRB 130831A.

The newly born magnetar may also provide a large energy input in the exploding progenitor, powering an energetic and luminous supernova explosion. This is in agreement with observations of Cano et al. (2014), who find a kinetic energy for the (non-relativistic) ejecta of SN 2013fu of $E_{\rm SN} = 1.9\times10^{52}$ erg and a peak absolute magnitude $M_V = -19.3$. We note that Greiner et al. 2015 found $E_{\rm SN} \simeq 10^{52}$ erg for SN2011~kl, a very bright supernova associated with GRB 111209A, whose properties can be explained by the energy injection of a newly born magnetar. In addition, the energetics of SN2013fu is quite typical of other SNe associated with GRBs (Cano et al. 2015).

As for the abrupt end of the X-ray emission of ``internal origin", with a very steep slope, we may attribute it to the delayed collapse of the magnetar into a black hole (Vietri \& Stella 1998, Lyons et al. 2010, Rowlinson et al. 2013). Once the magnetar has lost much of its rotational energy to power the jet, the weakened centrifugal forces may not be able to avoid the collapse. The timescale of such event, for a neutron star mass of $\sim 2 M_{\odot}$, initial period of $\sim 2\times10^{-3}$ s and magnetic field of $10^{15}$ Gauss would be $\sim 6\times10^4$~s (Vietri \& Stella 1998, their Eq. 1), comparable to the epoch of the steep drop in the X-ray light curve of GRB 130831A. Such a collapse should be relatively quick, and rapidly stop energy emission from the central object. If a magnetar collapsing into a black-hole is the right model, GRB~130831A might be a candidate for the production of Fast-Radio Bursts (FRBs), as described in Zhang (2014), as thus a target for observational campaigns in radio aimed at understanding the origin of FRBs.

% from $\alpha_X \simeq 0$ predicted by simple dipole spin-down process expected in young magnetars (which is relevant to some other X-ray afterglows)
 
%As mentioned above, the magnetar is thought to produce a roua ghly constant luminosity $L_0$ for a time $T_{em}$ because of the spin-down process, followed by a rapid decay $L \propto (t/T_{em})^{-2}$ afterwards. Another possibility to explain the sudden decrease of flux is that the magnetar was too massive to be stable and collapsed into a black hole. Its initially fast ration prevented the NS from collapsing into a black hole but, as the magnetar was losing rotational energy into outflow energy and perhaps gravitational waves, it rotation speed was decreasing until the centrifugal force was not able to halt the collapse.

%According to Zhang \& M\'esz\'aros (2002), the values of the parameters above are given by

%\begin{equation}\label{luminosity}
%    L_{0} = I \Omega_{0}^{2} \ 2 T_{em} \sim 10^{49} B^{2}_{15} P^{-4} _{-3} R^{6}
%    _{6} \, {\rm erg \, s}^{-1}\, ,
% \end{equation}

%where $I$, $B$, $P$, and $R$ are the momentum of inertia, the magnetic field, the period and radius of the magnetar, and 

%\begin{equation}\label{timescale}
%  T_{em} = \frac{ 3c^{2} I }{ B^{2} R^{6} \Omega_0^{2}} = 2.05 \times
%10^{3} (I_{45}
%  B^{-2}_{15} P^{2}_{-3} R^{-6}_{6} ) \, {\rm s}\, .
%\end{equation}

%By multiplying equations \ref{luminosity} and \ref{timescale}, we obtain

%\begin{equation}\label{eq2c}
% L_{0} T_{em} = 2.05 \times 10^{52} P^{-2}_{-3} \, {\rm erg} \, .
%\end{equation}

\subsubsection {A black hole with a fall-back accretion disk}
Another possibility we consider is that the central engine of this GRB could be a stellar black hole with a fall-back accretion disk (Kumar et al. 2008). Basically, in the supernova explosion associated with the GRB, the innermost part of the star collapses into a black hole. A continued fall-back of matter  - directly from the progenitor envelope or from supernova ejecta that failed to reach escape velocity - occurs at the centre. Some of this material does not accrete directly onto the black hole, but creates an accretion disk around it. Depending on the fall-back rate and the accretion time $t_{\rm acc}$ onto the black hole, the material of this disk can power relativistic ejecta from the black hole for a long time, which may produce both the prompt emission and a long-lived, slowly decaying X-ray flux.

  For the latter, Kumar et al. (2008) envisage a few possibilities. One is that the accreting disk has a low viscosity. It will thus take a long time, of the order of $\sim10^4-10^5$~s, for the all the disk material to accrete onto the black hole and power the jet. This model explains a long-lived plateau, but it predicts a flux decay with slope $\alpha \simeq 1.3$ at the end of this phase, which is not observed in GRB 130831A and other GRBs with similar features.
  
  Another possibility is that the disk has high viscosity. Then $t_{\rm acc}$ is much less than the fall-back time, and the emission basically traces the rate of the matter falling back onto the accretion disk. However, such a scenario cannot explain why the fall-back rate is less steep than expected: on theoretical grounds, we expect a fall-back rate to vary as $t^{-5/3}$ (Chevalier 1989). Secondly, the plateau slope and the sudden cut-off can be explained only by a particular density and angular momentum profile of the matter of the progenitor. A luminosity that evolves close to $t^{-1}$, as in the case of GRB 130831A, might be explained if the density profile of the stellar envelope has a profile of approximately $r^{-3}$, where $r$ is the radius from the centre of the star. A steep drop at the end of the plateau might be achieved if the material of the stellar envelope, which is the last to be accreted, has a relatively small angular momentum: this will cause the matter to fall rapidly onto the black hole and shut off the emission. Such a peculiar configuration, however, seems somewhat contrived and at odds with models of stellar progenitors (e.g. Woosley et al. 2011). Moreover, to support accretion for $\sim 10^5$~s, the disk should be unusually large and massive, leaving little mass for ejecta. Cano et al. (2014) find instead that the SN~2013fu, associated with GRB 130831A, has an ejecta mass of $M_{\rm ej} \approx 4.7$ $M_{\odot}$, which is typical of other GRB-SNe (e.g. Cano 2013). 
  
  For the model discussed, the jet luminosity is expected to be in the range of $10^{45}$ erg s$^{-1}$ at the end of the plateau, which is similar to the X-ray luminosity of the GRB 130831A afterglow at the end of the shallow decline phase. 

\subsubsection {A Binary Origin}
Barkov \& Komissarov (2010) conceive another scenario in which a black hole might power a GRB and long-lived outflow (see also Komissarov \& Barkov, 2009). If a compact object and a Wolf-Rayet (WR) star form a very close binary, such a system can go through a common envelope phase in which the compact object spirals in and can accrete the matter of the companion. The common envelope matter will have very high angular momentum and can take a long time to accrete. According to Barkov \& Kommisarov, the accretion timescale of such a system is:

\begin{equation}
t_d \simeq 8000 \left(\frac{\alpha}{0.01}\right)^{-1} \left(\frac{R_{\rm s}}{R_{\odot}}\right)^{3/2}  \left(\frac{M_{\rm c}}{2M_{\odot}}\right)^{2} \left(\frac{M_s}{10M_{\odot}}\right) ^{-7/2} {\rm s}
\label{Barkov}
\end{equation} 
where $\alpha$ represents the viscosity, $R_{\rm s}$ and $M_{\rm s}$ the radius and mass of the WR star, and $M_{\rm c}$ is the mass of the compact object. For massive compact objects, an accretion timescale of several tens of ks is not impossible\footnote{Note, though, that Eq. \ref{Barkov} is valid if $M_{\rm c}$ is considerably smaller than $M_{\rm s}$. However, if they become comparable, e.g. when we have a black hole of 10 $M_{\odot}$ and Wolf-Rayet star of similar mass, one has to find a different method to estimate $t_d$.}. According to Barkov \& Kommisarov (2010), during accretion the compact object will produce jets via the Blandford-Znajek  mechanism, and the jet luminosity will be in the range of $10^{49}$ erg s$^{-1}$, more than enough to explain the luminosity during the slow decline phase. Nonetheless, this scenario might suffer from the same problems as the previous one. If the viscosity of the accretion disc is low, which is required to keep the material around the black hole for $\sim1$~day, then the flux should not decrease quickly at the end of the plateau. The binary origin scenario could not reproduce the steep X-ray flux drop and/or it would require a peculiar structure of the WR star to explain the temporal dependence of the luminosity. We note, however, that in their papers, Barkov and Kommissarov do not discuss what happens at the end of the plateau, so we can only postulate.

Finally, we remark that this binary origin model can predict a supernova, like the one associated with GRB 130831A.

\subsubsection{Concluding remarks on the origin of the X-ray emission}
We have shown that the FS scenario (or any refreshed shock scenario) cannot explain the X-ray emission of GRB 130831A between the end of prompt emission and $\simeq 100$~ks, since such a model cannot entail the steep decay of the flux at that epoch. 

We have investigated whether such early X-ray emission can be attributed to dissipation processes occurring in the outflow of a newly born magnetar, produced via spin-down energy extraction of the compact object. We have found that the simplest model cannot explain the observations, because it predicts a flat X-ray light-curve followed by a steep drop. In the case of GRB 130831A, the decay slope before the break is $\alpha \simeq 0.8$, which is much steeper than what we expect in this model. However, a more elaborate model of the magnetar spin-down, in which the magnetic field is expected to decay as the rotation time increases, predicts a luminosity decay more consistent with observations and a duration that can extend up to tens of ks. Moreover, the anticipated X-ray luminosity is in the right range. The assumed initial parameters - initial period $P$ and magnetic field $B$ - for the newly born magnetar that would produce such a X-ray light-curve are $P = 1 - 2$~ms and $B\simeq10^{15}$ G, which are expected on theoretical grounds for such an object.

 We have also discussed whether the compact object could be a black hole rather than a magnetar. In order to power emission for such a long time, the black hole should be surrounded by an extended disk, perhaps produced by fall-back material of the supernova associated with the GRB. While this model still may predict the right luminosity and duration for the ``internal" X-ray emission, it would require an anomalous distribution of angular momentum in the progenitor star, and peculiar fall-back rates. Similar advantages and disadvantages may be present in a model in which the compact object forms a binary with a WR star and spirals in, blowing up the star into a massive disk. All in all, we deem the magnetar model with a decaying magnetic field the most plausible of those presented so far to explain the properties of GRB 130831A.

\subsection{{\bf Energy budget of the X-ray radiation between the end of the prompt emission and the 100 ks drop.}}
%One can easily calculate the energy produced, between 0.3 and 10 keV and the beginning of observations and the 100~ks break time.
In the cosmological rest frame, the X-ray internal emission begins no later than $200/(1+0.479) = 135$~s and lasts until $\simeq 98.2\times10^3 / (1+0.479) \simeq 66.4\times10^3$~s. Taking into account cosmological corrections, the 0.3--10 keV luminosity at 135 s is $2.1\times10^{47}$~erg s$^{-1}$; we assume that from this epoch the luminosity decreases as $t^{-0.8}$ up to 66.4~ks. We subtract the amount of X-ray emission produced by the FS (see section \ref{early}), and we obtain a total energy $E_{\rm X} \simeq 2.8 \times 10^{50}$ erg. Such a value represents $\simeq 2.5\%$ of the energy emitted during the prompt phase, and only $\simeq 0.25\%$ of the kinetic energy of the relativistic ejecta producing the FS emission. We note, however, that the internal emission may extend below 0.3 keV and above 10 keV. Thus, $E_{\rm X}$ and the percentage we have determined represent a lower limit.\\

\subsection{Energy breakdown of GRB 130831A and the associated supernova}

 We know that the kinetic energy of the ejecta of SN 2013fu, the supernova associated with GRB 130831A, is $\simeq 1.9\times10^{52}$ erg (Cano et al. 2014). This is a factor $\sim 6$ lower than the kinetic energy of the relativistic ejecta and comparable to the energy emitted during the prompt emission, and $\sim 60$ times higher than the energy associated with the X-ray emission of ``internal origin". We do caution, however, that the estimated SN kinetic energy is {\it isotropic}, while the energy of the relativistic ejecta, the prompt and the early X-ray energetics must be corrected for an unknown beaming factor. GRB 130831A does not show the signature of a jet break (Sari, Piran \& Halpern 1999) within our observations, so we cannot derive the beaming angle $\theta_j$ of the ejecta and thus the beaming factor. However, we can set limits. The detection of the X-ray afterglow by {\it Chandra} at 1430 ks after the trigger indicates that the FS X-ray afterglow had a decay slope of $\simeq 1$ up to that epoch and no jet break had yet occurred.\\
Following Zhang et al. 2009, the beaming angle of the ejecta in a medium with constant density can be estimated as

\begin{equation}
\theta_{jet} = 0.12 \left(\frac{t_{jet,d}}{1+z}\right)^{3/8} \left(\frac{E_{K,53}}{n}\right)^{-1/8}\, \rm{rad}
\end{equation} 
 
\noindent where $t_{jet,d}$ is the jet break time in days. Our first {\it Chandra} observation took place 16.6 days after the trigger and we adopt the values $E_K$ and $n$ determined from our modeling; we note that the exact value of $\theta_j$ depends only weakly on the value of these parameters anyway. We infer that $\theta_j \gtsim 0.123$~rad; thus the lower limit on the beaming factor is $f_b \simeq \theta_j ^2 / 2 \simeq 7.56\times10^{-3}$. By definition, the maximum value for the beaming factor is $f_b=1$ when the source is isotropic.\\
If the beaming factor of GRB 130831A emission were $f_b = 7.56\times10^{-3}$, the total energy budget of this event and its supernova would be $\simeq 2.0\times10^{52}$ erg. The prompt energy $E_{\gamma}$, the energy emitted in X-rays up to 100 ks $E_{\rm X}$, and the kinetic energy $E_{\rm K}$ of the relativistic ejecta would be $\simeq 0.4\%$, $\simeq 0.01\%$ and $\simeq 4.3\%$ of the total energy budget. If, as an extreme and unlikely case, GRB 130831A were isotropic, its total energy budget would be $1.5\times10^{53}$ erg; the above percentages would become $\simeq 7\%$, $\simeq 0.2\%$ and $\simeq 80\%$. The kinetic energy of the relativistic ejecta is {\it at least} $\simeq 4.3\%$ of the total energy produced by the GRB and the SN. Moreover, the fraction of energy going into the ``internal emission" X-rays is always rather small, being substantially less than $1\%$ in both cases.\\
If GRB 130831A and its SN are powered by a magnetar, the total energy budget cannot be $\gtsim 3\times10^{52}$ erg (see section \ref{origin}). To not exceed this limit, the beaming factor of the GRB must be $f_b \ltsim 0.1$. If $f_b=0.1$, $E_{\gamma}$, $E_{\rm X}$, and $E_{\rm K}$ represent $\simeq 3.3\%$, $\simeq 0.1\%$ and $37\%$ of the total energy. In reality, we should expect these percentages to be between those of the $f_b=0.1$ and~$f_b=7.56\times10^{-3}$ cases if GRB 130831A is actually powered by a magnetar. The breakdown is presented again in Table \ref{breakdown}.

\subsection{Early Afterglow}\label{early}

In our analysis, we have focused on the afterglow emission between 15 and 230 ks. It is worth exploring whether our model can explain the interesting features of the early afterglow, especially the optical band.

The initial flare peaks at 730~s. It takes place both in the X-ray and optical bands, but it is very pronounced in the latter. Its rapid temporal evolution (the optical flux increased by a factor of $\sim5$ between 400 and 800 s after the trigger) suggests that it could be explained in the context of internal dissipation processes which occurred in the outflow, when the Lorentz factor is very high and relativistic effects cause rapid variations of the observed flux. Thus, GRB 130831A may show a clear example of internal dissipation that produces strong emission in the optical other than in the X-ray. Alternatively, the optical flare might flag the onset of FS emission. However, if this were the case, the decay slope after the flare peak would be consistent with the decay slope of the late optical afterglow $\alpha_{\rm opt}$. Instead, the decay rate after the flare peak is $\alpha=1.79\pm0.02$, which is inconsistent with $\alpha_{\rm opt}=1.59\pm0.03$ found later.
After the flare, the early optical emission shows a plateau up to a few ks (see Section \ref{LCSEDS}). Typically, an early slow optical decay is interpreted as energy injection, which ends at the time of the break. However, this interpretation might be difficult in our scenario, because the energy injection would have to stop at $\sim 5$ ks while, according to our analysis, the GRB outflow is still active at $\simeq 100$ ks.

A possibility is that the optical plateau basically results from the combination of the decaying optical flare and the rising of the FS peak, which has been shown for a number of {\it Swift} bursts for which early optical light-curves are available (Oates et al. 2009). The peak Lorentz factor $\Gamma$ of the ejecta then can be calculated (Molinari et al. 2007 and references therein) as
\begin{equation}\label{eq3}
\Gamma = 160 \left( \frac{E_{\gamma,53} (1+z)^3 } {\eta_{0.2}~n~t_{dec,2}^3} \right)^{1/8}
\end{equation}
For a deceleration time $t_{\mathrm dec} = 4000$~s and adopting the values of energy and density we have determined above, the resulting peak Lorentz factor would be $\Gamma \simeq 100$, which is within the overall distribution of Lorentz factors for GRB afterglows (Oates et al. 2009). The reason for such a late deceleration, in our model, comes naturally given the low density of the circumburst medium.\\

%We address this issue in the following.

%The late onset of FS emission eases another potential problem of our model.

The late emission, which we attribute to FS in our modeling, seems to have a relatively steep decay slope. So one may wonder whether it could give some important contribution to the X-ray flux as well at an earlier epoch. If we extrapolate the late X-ray flux to earlier epochs using a decay slope of $\simeq 1.6$, it would become comparable to or even higher than the observed X-ray flux at the end of the first orbit (at $\simeq 800$~s) and the shape of the X-ray light-curve would differ from what we see. However, if the FS onset occurs at $\simeq 4000$~s, this problem is avoided. As for the observations from $\simeq 9$~ks (i.e. the beginning of the second orbit) onwards, at 10 ks the flux by FS emission is $\simeq 6 \times10^{-14}$ erg cm$^{-2}$ s$^{-1}$ $\times (\frac{10}{173})^{-1.59} = 5.5\times10^{-12}$ erg cm$^{-2}$ s$^{-1}$, which is $\sim 3$ times weaker than the observed X-ray flux. Afterwards, the FS flux decreases faster than that of ``internal origin" that we see, which decays with a slope of $\simeq 0.8$. Similarly, for a $\beta_{\rm OX} = 1.03$ and flux density in the R band ($4.6\times10^{14}$~Hz) of $\simeq 410$ $\mu$Jy at 10~ks, the expected X-ray 0.3-10 keV flux is $\simeq 5.4\times10^{-12}$ erg cm$^{-2}$ s$^{-1}$. This is again $\sim 3$ times lower than the X-ray flux observed at 10 ks. Thus, the X-ray flux from 9 ks up to the steep drop is not predominantly produced by the FS.
%We cannot exclude, however, that the FS contributes to some extent to the X-ray flux a few thousands seconds after the trigger.

\section{Conclusions}
 We have discussed the case of the long {\it Swift} GRB 130831A. The X-ray afterglow of this burst initially shows a shallow decay. However, at $\simeq 100$ ks the X-ray light-curve breaks to an unusually steep decay slope $\simeq 6$, which cannot be explained by the standard FS model. Late XRT and especially {\it Chandra} observations show that the X-ray afterglow has a successive break to a more sedate decay with a slope $\simeq 1.1$.
 
 The well-sampled optical afterglow shows no change of slope concurrent with the steep break in the X-ray band, which we interpret as arising from a different mechanism. By means of data taken by {\it Swift}, {\it Chandra} and several ground-based observatories, we have shown that both the optical and late X-ray emissions, after $\simeq 200$~ks after the trigger, have the typical decay and spectral slopes of GRB afterglows explained by the FS model.
  
 We have interpreted the X-ray and optical afterglow as the superposition of two emission components. One component is of ``internal" origin, generated within a relativistic outflow and responsible for the early X-ray emission up to $\simeq 100$~ks; the outflow is produced by either the spin-down of a newly formed magnetar or a black hole feeding on fall-back matter. A second component, responsible for the late X-ray and optical from a few ks after the trigger, is the typical FS emission. When the magnetar has lost much of its rotational energy or the black hole does not accrete and does not power the outflow any longer, the first component dies off and we see the steep X-ray decay that lasts until the standard FS emission emerges. We believe that the magnetar model is favoured, since the other scenario would require a rather peculiar stellar progenitor structure and fall-back process. 

 Modeling the late optical and X-ray afterglow, we have inferred the kinetic energy of the relativistic ejecta and thus an efficiency $\eta\simeq 0.07$ of the ``central engine" of this GRB to produce $\gamma$-ray emission. This efficiency is smaller than that of other bursts that show emission of internal origin (L\"u \& Zhang 2014; although these were examined in the X-ray band only), and more typical of those GRBs in which no internal emission is clearly visible. Thus, GRB 130831A may represent a ``trait d'union" between GRBs with different dominant emission processes. 
 
More importantly, gathering the information on the kinetic energy of the supernova associated with 130831A, we have provided a breakdown of the energetics of the GRB and its associated supernova. We have found that, regardless of the nature of the central engine and unknown collimation of the ejecta, at least $\simeq 4.3\%$ of the total energy of the event is coupled with relativistic ejecta; and less (probably significantly less) than $\simeq0.2\%$ of the energy goes into X-ray emission of ``internal origin" lasting up to 100 ks in our case; this component produces a factor $\sim 30$ less energy than that released in $\gamma$-ray during the prompt emission.\\

  Showing several emission processes at work, GRB~130831A has offered us the opportunity to investigate the complete phenomenon of a supernova with a central engine that produces the explosion and drives an energetic relativistic outflow, where dissipation processes take place for a long time.

% Thus, GRB 130831A may represent a ``trait d'union" between diverse GRBs, and help us understand the different properties that central engines of GRBs may have. 

\section*{Acknowledgments}

We thank H. Tananbaum for granting us DDT observations of GRB 130831A with {\it Chandra}. This research has made use of data obtained from the Chandra Data Archive and the Chandra Source Catalog, and software provided by the Chandra X-ray Center (CXC) in the application packages CIAO, ChIPS, and Sherpa.

MDP, MJP, SRO and AAB thank UK Space Agency for financial support. This work made use of data supplied by the UK Swift Science Data Centre at the University of Leicester.

AP, AV acknowledge partial support by RFBR grants 12-02-01336, 13-01-92204, 14-02-10015 and 15-02-10203.

AJCT and SRO acknowledges support from the Spanish Ministry Grant AYA 2012-39727-C03-01

S{\bf S} acknowledges support from CONICYT-Chile FONDECYT 3140534, Basal-CATA PFB-06/2007, and Project IC120009 "Millennium Institute of Astrophysics (MAS)" of Iniciativa Cient\'{\i}fica Milenio del Ministerio de Econom\'{\i}a, Fomento y Turismo.

DAK acknowledges financial support by the  Th\"uringer Landessternwarte Tautenburg, and the Max-Planck Institut  f\"ur Extraterrestrische Physik.

ZC gratefully acknowledges support by a Project Grant from the Icelandic Research Fund.

We thank the RATIR project team and the staff of the Observatorio Astron�mico Nacional on Sierra San Pedro M\'{a}rtir. RATIR is a collaboration between the University of California, the Universidad Nacional Auton\'{o}ma de M\'{e}xico, NASA Goddard Space Flight Center, and Arizona State University, benefiting from the loan of an H2RG detector and hardware and software support from Teledyne Scientific and Imaging. RATIR, the automation of the Harold L. Johnson Telescope of the Observatorio Astron�mico Nacional on Sierra San Pedro M�rtir, and the operation of both are funded through NASA grants NNX09AH71G, NNX09AT02G, NNX10AI27G, and NNX12AE66G, CONACyT grants INFR-2009-01-122785 and CB-2008-101958 , UNAM PAPIIT grant IN113810, and UC MEXUS-CONACyT grant CN 09-283.

Partly  based  on  observations  carried  out with  the  10.4m  Gran Telescopio Canarias  (GTC) installed  in the Spanish  Observatorio del Roque  de  los Muchachos  of  the  Instituto  de Astrof\'{\i}sica  de Canarias in the island of La Palma.

%IRAF is distributed by the National Optical Astronomy Observatory, which is operated by the Association of Universities for Research in Astronomy, Inc., under cooperative agreement with the National Science Foundation.

\clearpage

%\clearpage

%\clearpage

%\clearpage

\begin{table*}
\begin{center}
\caption{Time resolved spectral analysis. ST = settling exposure, WT = Window Timing Exposure, PC = Photon Counting exposure. Note: errors are at 90\% C.L.  \label{tab:xrtspec}}
\begin{tabular}{llllllllllll}
\hline
Name & Start time & End Time & excess N$_\textrm{H}$  & $\beta$ & Observed flux & Unabsorbed flux & Cstat & dof \\
            &       (s)          &        (s)         & $(10^{20}$cm$^{-2})$      &                       & ($\times$ erg cm$^{-2}$ s$^{-1}$) & ($\times$ erg cm$^{-2}$ s$^{-1}$) & & \\
\hline
& & & & & & & & \\
ST   & 115.8           & 125.2     	 & $<700$          		& 1.7$\pm$0.2 		& 7.93 $\times10^{-10}$     & 1.06 $\times 10^{-9}$       &  125.4    & 146\\
WT1  & 132.0         & 157.9     	 & $<400$          		& 1.7$\pm$0.1 		& 6.91 $\times10^{-10}$     & 9.21 $\times 10^{-10}$     &  163.3   & 159\\
WT2  & 157.9         & 205.6     	 & $<500$         		& 1.3$\pm$0.1 		& 3.34 $\times10^{-10}$     & 4.06 $\times 10^{-10}$     &  182.2     & 184\\
PC1  & 207.2          & 287.2     	 & $<500$       	       & 1.2$\pm$0.3 		& 1.70 $\times10^{-10}$     & 2.03 $\times 10^{-10}$     &  82.5     & 76 \\
PC2  & 287.2          & 367.2     	 & $<100$         		& 0.7$\pm$0.3 		& 1.47 $\times10^{-10}$     & 1.61 $\times 10^{-10}$     &  60.3     & 60 \\
PC3  & 367.2          & 437.2     	 & $<20$        		& 1.1$\pm$0.4 		& 9.69 $\times10^{-11}$     & 1.14 $\times 10^{-10}$     &  37.1     & 41 \\
PC4  & 437.2          & 847.2     	 & $<900$       		& 0.8$\pm$0.1 		& 1.47 $\times10^{-10}$     & 1.63 $\times 10^{-10}$     &  199.4   & 216\\
PC5  & 9891.6        & 132418.7  & 6.8$^{+3.3}_{-3.1}$    & 0.77$\pm$0.12     & 6.23 $\times10^{-11}$    & 6.55 $\times 10^{-11}$     &  305.4   & 367\\
PC6  & 171385.0   & 1193789.0    & $<700$                       & 1.0$\pm$0.9         & 1.52 $\times10^{-14}$    & 1.80 $\times 10^{-14}$     &  10.3      & 17 \\
\hline
\end{tabular}
\end{center}
\end{table*}

\begin{table*}
\begin{center}
\caption{Results of the temporal analysis of the X-ray emission of GRB 130831A, which includes XRT and {\it Chandra} data. We show the results of fitting the data with two models. The first model, {\it BPLs + flare}, consists of power-law plus broken power-law plus power-law plus gaussian flare and it fits the whole X-ray dataset. The second model, {\it 2powls}, includes 2 power-law components and it fits the data after 100~ks. We show the decay indices, break time and the centre (GC), width (GW) and flux normalisation (GN) of the gaussian flare, of the first model. The latter model has $\alpha_{X,3}$ and $\alpha_{X,4}$ only.}
\label{tab: xrt time analysis}

\setlength{\tabcolsep}{3pt}
\begin{tabular}{lcc ccc ccc cc}
\hline
Model & $\alpha_{X,1}$      & $\alpha_{X,2}$  & $t_{b}$         & $\alpha_{X,3}$  & GC & GW & GN  & {\bf $\alpha_{X,4}$} & $\chi^2/dof$ \\
            &                            & (ks)                       &          (ks)                  &                     &           (ks)                  &     (ks)  &    ($\times10^{-11}$ erg cm$^{-2}$ s$^{-1}$)     &       &                                    &   \\                    
\hline\vspace{1mm}
BPLs + flare & $5.97 \pm 0.01$ & $0.80^{+0.01} _{-0.02}$ & $98.26^{+2.94} _{-3.30}$ & $5.9^{+1.0} _{-0.4}$ & $0.73\pm0.03$ & $0.11\pm0.02$ & $11.2\pm1.6$ & $0.90^{+0.11} _{-0.05}$ & $50.7/48$\\

2powls & & & & $6.8^{+2.0} _{-1.5}$ & & & & $1.11 ^{+0.22} _{-0.29} $ & 2.4/3\\

\hline
\end{tabular}
\end{center}
\end{table*}

\begin{table*}
\begin{center}
\caption{Photometry of the Afterglow of GRB 130831A. All magnitudes are in the Vega system. The full table is available online.}
\label{Photometry}
\begin{tabular}{llllllllllllll}
$t-t_0$ 	    & Exposure time 	    &              mag        	      & Filter      & Telescope\\ \hline \vspace{1mm}
      (s)        &          (s)                 &                                          &               &  \\ \vspace{1mm}
& & & & \\

484		&	10	  &  $		14.86	^{+0.09} _{-0.08}	  $ &	$uvw2$	&	UVOT	\\ \vspace{1mm}
633		&	10	  &  $		14.03	^{+0.06} _{-0.05}	  $ &	$uvw2$	&	UVOT	\\ \vspace{1mm}
783		&	10	  &  $		13.67	\pm	0.05		  $ &	$uvw2$	&	UVOT	\\ \vspace{1mm}
10346	&	450	  &  $		17.53	^{+0.05} _{-0.04} 	  $ &	$uvw2$	&	UVOT	\\ \vspace{1mm}
43951	&	2844	  &  $		19.44	^{+0.15} _{-0.13} 	  $ &	$uvw2$	&	UVOT	\\ \vspace{1mm}
123835	&	3414	  &  $	      >21.27			        	  $ &	$uvw2$	&	UVOT	\\ \vspace{1mm}
... 			&   ... 		  &			...						     &       ...			&     ...  \\
\hline
\end{tabular}
\label{UVOIR}
%\end{center}
\end{center}
\end{table*}

\begin{table*}
\begin{center}
\caption{Calibration stars which were used for the optical telescopes. RATIR used a much longer list of calibration stars, but we list only those in common with other instruments.}
\label{tab:referencestars}
\begin{tabular}{llll}
Catalogue & RA   &     Dec	&  Telescopes \\ 
\hline
APASS & 358.594169 & +29.341196  &  Skynet RATIR \\
APASS & 358.660796 & +29.429860  &  Skynet RATIR \\
APASS & 358.548308 & +29.427729  &  Skynet RATIR \\

\hline
%SDSS & J235443.71+292539.3 & ISON RATIR  \\
%SDSS & J235423.64+292452.8 & ISON RATIR \\ 
%SDSS & J235434.33+292716.7 & ISON RATIR  \\ 
SDSS & 358.682125  & +29.427583 & IKI RATIR  \\
SDSS & 358.660833  & +29.429806 & IKI \\
SDSS & 358.598500  & +29.414667 & IKI RATIR \\ 
SDSS & 358.643042  & +29.454639 & IKI RATIR  \\
SDSS  &	358.63521  &  +29.42802 &  NOT, LT, RATIR \\
SDSS  & 358.63296  &  +29.42655 &  NOT, LT, RATIR \\
SDSS  &	358.62188  &  +29.42342 &  NOT, LT, RATIR \\
SDSS  &	358.63986  &  +29.41882 &  NOT, LT, RATIR \\
SDSS  &	358.64747  &  +29.41695 &  NOT, LT, RATIR \\
\hline 
\end{tabular}
\end{center}
\end{table*}

\begin{table}
\begin{center}
\caption{Flux densities corresponding to zero magnitudes used in the conversion of magnitudes to flux densities for the light-curves shown in Fig.~2.}
\label{tab:fluxconversions}
\begin{tabular}{cc}
Filter & Flux (Jy) \\ 
\hline
ISON unfiltered& 2786 \\
UVOT   $u$     & 1445 \\
UVOT   $UVW1$  &  888 \\
UVOT   $UVM2$  &  769 \\
UVOT   $UVW2$  &  735 \\
Skynet $B$     & 4127 \\
Skynet $V$     & 3690 \\
Skynet $R$     & 3103 \\
Skynet $I$     & 2431 \\
Skynet $g'$    &  363 \\
RATIR $r'$     & 3147 \\
RATIR $i'$     & 2590 \\
RATIR $Z$      & 2211 \\
RATIR $Y$      & 2040 \\
RATIR $J$      & 1564 \\
RATIR $H$      & 1007 \\
\hline 
\end{tabular}
\end{center}
\end{table}

\begin{table*}
\begin{center}
\caption{Results of the temporal analysis of the optical emission of GRB 130831A between 3.5 ks and 15 ks (upper part) and 15 ks and 13 Ms (lower part). Time is expressed in ks, while the constant flux is in $\mu$Jy. Since the host galaxy constant flux is not important during the first fit, which does not extend up to late times anyway, we have omitted it.}
\label{tab: opt time analysis}
\begin{tabular}{llllllll}
\hline
Filter & $\alpha_1$ & $t_{b,1}$  & $\alpha_2$          & Const          & $\chi^2/dof$ & & \\
          &                      &     (ks)			  &			      &	($\mu$Jy) & 	 			& & \\
\hline
& & & & & & & \\
No Filter   & $0.06^{+0.19} _{-0.20}$ & $4.79^{+0.16} _{-0.17}$ & $1.62\pm 0.07$ & -- & $86/53$ \\
$B$   	& $0.42 \pm 0.08$ 		& $4.90\pm0.08$ & $1.45\pm 0.03$ & -- & $81/49$  \\
$V$   	& $0.86\pm0.03$ & $6.47\pm0.10$ & $1.65\pm0.03$ & -- &$160/57$   \\
$R$   	& $0.64^{+0.04} _{-0.06}$ & $5.56^{+0.11} _{-0.16}$ & $1.56\pm0.02$ & -- & $232/57 $ \\
$I$     	& $0.92\pm0.03$ & $6.99^{+0.17} _{-0.19}$ & $1.82^{+0.05} _{-0.06}$ &-- &$119/50$   \\
     &             &             &            &  &       \\
\hline
& & & & & & & \\
%Power-law + constant & & & & \\
$R$ & $1.60\pm 0.03$ & -- & -- & $0.81\pm0.14$   & 9.4/10   \\
$r'$  & $1.49\pm0.06$  & -- & -- & $0.96\pm0.11$  & 30.7/14 \\                    
$i'$  & $1.64\pm0.07$  & -- & -- & $0.73\pm0.08$  & 29.2/15 \\
\hline
\end{tabular}
\end{center}
\end{table*}

\begin{table*}
\begin{center}
\caption{Best fit parameters obtained when fitting the 173 ks (2 days) SED with a single broken power-law and Milky Way extinction law. We indicate the reddening in our Galaxy and that in the host at $z=0.479$ separately.
The absorption in our Galaxy (z=0) and in the host of the GRB (z=0.479) has been fixed to the best fit value
of the X-ray data.} 
\label{SEDfit}
\begin{tabular}{llllll}
\hline
N$_H$ at z=0  & N$_H$ at z=0.479 & $E(B-V)$ at z=0    & $E(B-V)$ at z=0.479   &$\beta_{\rm OX}$  & $\chi^2/\mathrm{dof}$ \\
                         &	$\times10^{22}$				 &		(mag)	    &		(mag)			 &					&						\\
\hline
& & & & & \\
$4.8\times10^{20}$ & $0.068$ & 4.0$\times10^{-2}$ & $1.8\pm1.3\times10^{-2}$ & $1.03^{+0.05} _{-0.04}$    & 12.7/9 \\

\hline
\end{tabular}
\end{center}
\end{table*}

\begin{table*}
\begin{center}
\caption{Breakdown of energetics of GRB 130831A and its associated SN 2013fu into energy emitted in $\gamma$-rays $E_{\gamma}$, energy produced in X-rays of internal origin $E_{\rm X}$, and kinetic energy associated with the relativistic GRB ejecta $E_{\rm K}$. These values are corrected for beaming corresponding to the beaming factor $f_b$.
The kinetic energy of the SN is $E_{\rm SN} = 1.9\times10^{52}$ erg (Cano et al. 2014), and the total energy is $E_{\rm tot} = E_{\rm SN} + E_{\gamma} + E_{\rm X} + E_{\rm K}$.}
\label{breakdown}
\begin{tabular}{llllllll}
\hline
Beaming factor $f_b$ & $E_{{\rm tot},52}$ & $E_{\gamma,{\rm corr}}$ & $E_{\rm X}$ & $E_{\rm K}$ & \\
\hline
                                    &         &	       &	  &		     & \\
1 (isotropic)              					& 14.8    & 7.2\%          & 0.19\%   & 80\% \\
0.1 (magnetar limit) 					& 3.2      & 3.3\%                  & 0.09\%   & 37\% \\
$7.56\times10^{-3}$ (lower limit)    & 2.0       & 0.39\%         & 0.01\%          & 4.3\% \\
\hline
\end{tabular}
\end{center}
\end{table*}

\label{lastpage}


\begin{thebibliography}{99}

\bibitem[\protect\citeauthoryear{Ahn}{2012}]{2012ApJS..203...21A} Ahn C.P, Alendandroff, R., Allende-Prieto, C., 2012, ApJS, 203, 21

\bibitem[\protect\citeauthoryear{Amati}{2006}]{2006MNRAS...372..233A} Amati L., 2006, MNRAS, 372, 233

\bibitem[\protect\citeauthoryear{Amati et al.}{2009}]{2009A&A...508..173A} Amati L., Frontera F., Guidorzi C. 2009, A\&A, 508, 173

\bibitem[\protect\citeauthoryear{Antoniadis et al.}{2013}]{ant2013} Antoniadis J., Freire P., Wex C. 2013, Science, 508, 173

\bibitem[\protect\citeauthoryear{Arnaud et al.}{1996}]{arn96} Arnaud K. A., 1996, Astronomical Data Analysis Software and Systems V, eds. Jacoby G. and Barnes J., p17, ASP Conf. Series volume 101. 

\bibitem[\protect\citeauthoryear{Band et al.}{1993}]{1993ApJ..413..281G} Band D., Matteson J., Ford L., et al., 1993, \apj, 413, 281

\bibitem[\protect\citeauthoryear{Barkov \& Komissarov et al.}{2010}]{2010MNRAS.401.1644B} Barkov M., \& Komissarov S., 2010, MNRAS, 401, 1644

\bibitem[\protect\citeauthoryear{Barthelmy et al.}{2005}]{2005SSRv..120..143B} Barthelmy S.~D., Barbier L.~M., Cummings J.~R., et al., 2005, SSR, 120, 143

\bibitem[\protect\citeauthoryear{Bernardini et al.}{2014}]{2014MNRAS.439L..80B} Bernardini M. G, Campana S., Ghisellini G., et al., 2014, MNRAS, 439L, 80

\bibitem[\protect\citeauthoryear{Bertin \& Arnouts}{1996}]{1996A&AS..117..393B} Bertin E., Arnouts S., 1996, A\&A, 117, 393

\bibitem[\protect\citeauthoryear{Bertin}{2010}]{2010ascl.soft10068B} Bertin E., et al., 2010, Astrophysics Source Code Library

\bibitem[\protect\citeauthoryear{Blandford \& Znajek}{1977}]{1977MNRAS.179..433B} Blandford R. D. \& Znajek R. L., 1977, MNRAS, 179, 433

\bibitem[\protect\citeauthoryear{Bloom et al.}{2001}]{2001AJ....121.2879B} Bloom J.~S., Frail D.~A., Sari R. 2001, AJ, 121, 2879

\bibitem[\protect\citeauthoryear{Burrows et al.}{2005}]{2005SSRv..120..165B} Burrows D.~N., Hill J.~E., Nousek J.~A., et al., 2005, SSR, 120, 165

\bibitem[\protect\citeauthoryear{Butler et al.}{2012}]{2012SPIE.8446E.10B} Butler N., Klein C., Fox O., et al., 2012, SPIE, 8446, 10

\bibitem[\protect\citeauthoryear{Butler et al.}{2013}]{Butler2013} Butler N., Watson A., Kutyrev A., et al., 2013, GCN Circ. 15165

\bibitem[\protect\citeauthoryear{Breeveld et al.}{2011}]{2011AIPC.1358..373B} Breeveld A.~A., Landsman W., Holland S.~T., et al., 2011, AIPC, 1358, 373

\bibitem[\protect\citeauthoryear{Cano et al.}{2014}]{2014AA} Cano Z., De Ugarte Postigo A., Pozanenko A., et al.,  2014, A\&A, 568, 19

\bibitem[\protect\citeauthoryear{Cano}{2013}]{2013MNRAS.434.1098C} Cano Z., 2013, MNRAS, 434, 1098

\bibitem[\protect\citeauthoryear{Cenko et al.}{2011}]{2011ApJ...732..29C} Cenko S. B., Bloom J. S., Kulkarni S. R., et al., 2011, MNRAS, 420, 2684

\bibitem[\protect\citeauthoryear{Chandra}{2012}]{chandra2012} Chandra P. \& Frail, D. A. \apj, 746, 156

\bibitem[\protect\citeauthoryear{Cepa et al.}{2000}]{cepa2000} Cepa J., Aguiar, M., Escalera, V. G., et al., 2000, SPIE 4008, 623

\bibitem[\protect\citeauthoryear{Chevalier}{1989}]{1989ApJ...346..847C} Chevalier R.~A., 1989, \apj, 346, 847

\bibitem[\protect\citeauthoryear{Curran et al.}{2010}]{2010ApJ...716L.135C} Curran P. A., Evans P. A., De Pasquale M., et al., ApJL, 716, 135

\bibitem[\protect\citeauthoryear{Cucchiara et al.}{2013}]{cu2013} Cucchiara A., Perley D., 2013, GCN Circ. 15144

\bibitem[\protect\citeauthoryear{Dall'Osso et al.}{2011}]{2011A&A....526.121D} Dall'Osso S., Stratta G., Guetta D., et al., 2011, A\&A, 526 121

\bibitem[\protect\citeauthoryear{De Ugarte-Postigo}{2012}]{dup2012} De Ugarte Postigo A., Lundgren A., Martin S., et al. A\&A, 538, 2012

\bibitem[\protect\citeauthoryear{Duffel \& MacFadyen}{2014}]{2014ApJ...791L...1D} Duffel P. C. \& MacFadyen A.  2014, ApJL, 791, 1

\bibitem[\protect\citeauthoryear{Evans et al.}{2009}]{2009MNRAS...399..1167C} Evans P. A., Beardmore A. P, Osborne J. P., et al., 2009, MNRAS, 399, 1167

\bibitem[\protect\citeauthoryear{Gao et al.}{2013}]{2013NewAR..57..141G} Gao, H., Lei, W.-H., Zou, Y-.C. et al., 2013, NewAR, 57, 141

\bibitem[\protect\citeauthoryear{Gehrels}{1986}]{1986ApJ...303..336G} Gehrels N., 1986, \apj, 303, 336

\bibitem[\protect\citeauthoryear{Gehrels et al.}{2004}]{2004ApJ...611.1005G} Gehrels N., Chincarini G., Giommi P., et al., 2004, \apj, 611, 1005

\bibitem[\protect\citeauthoryear{Golenetskii et al.}{2013}]{2004ApJ...611.1005G} Golenetskii S., Aptekar R., Fredericks D., et al., 2013, GCN Circ., 15145

\bibitem[\protect\citeauthoryear{Granot}{2006}]{2007RMxAC..27..140G} Granot J., 2006, RMXAC, 270, 140.

\bibitem[\protect\citeauthoryear{Greiner et al.}{2015}]{2004ApJ...611.1005G} Greiner J., Mazzali, P. A., Kann, D. A., et al., 2015, Nature, 523, 189

\bibitem[\protect\citeauthoryear{Falcone et al.}{2007}]{2007ApJ...671.1921F} Falcone A.~D., Morris D., Racusin J.~L., et al., 2007, \apj, 671, 1921

\bibitem[\protect\citeauthoryear{Fruscione et al.}{2006}]{2006SPIE.6270E..60F} Fruscione A., et al. 2006, SPIE proc., 6270

\bibitem[\protect\citeauthoryear{Henden \& Munari}{2014}]{2014CoSka..43..518H}  Henden A. \& Munari U., 2014, Contributions of the Astronomical Observatory Skalnat\/e Pleso, 43, 518

\bibitem[\protect\citeauthoryear{Jarosik et al.}{2011}]{2011ApJS..192...14J} Jarosik N., Bennet, C. L., Dunkley J., et al., 2011, ApJS, 192, 14.

\bibitem[\protect\citeauthoryear{Hjorth et al.}{2012}]{2012ApJ...756..187H} Hjorth J., Malesani D., Jakobsson P., et al., 2012, \apj, 756, 187

\bibitem[\protect\citeauthoryear{Jordi et al.}{2016}]{2006A&A...460..339J} Jordi K., Grebel E.~K., Ammon K., 2006, 460, 339

\bibitem[\protect\citeauthoryear{Kalberla et al.}{2005}]{2005A&A...440.775K} Kalberla, P.~M.~W., Burton W.~B, Hartmann D., et al., 2005, \apj, 440, 775

\bibitem[\protect\citeauthoryear{Klein et al.}{2012}]{2012SPIE.8543.2S} Klein, C.~R., Kubanek P., Butler N.~R., et al., 2012, 8453, 2

\bibitem[\protect\citeauthoryear{Klose et al.}{2013}]{2013GCN...15320.1T} Klose S., Nicuesa Guelbenzu A., Kr\"uhler T., et al., 2013, CBET 3677

\bibitem[\protect\citeauthoryear{Kobayashi \& Zhang}{2007}]{2007ApJ...655..973K} Kobayashi S. \& Zhang B., 2007, \apj, 655, 973

\bibitem[\protect\citeauthoryear{Komissarov \& Barkov}{2009}]{2009MNRAS.397.1153K} Komissarov S., Barkov M., 2009, MNRAS, 397, 1153

\bibitem[\protect\citeauthoryear{Kraft et al.}{1991}]{1991ApJ...374..344K} Kraft, R.~P., Burrows D.~N., Nousek J.~A. 1991, \apj, 374, 344 

\bibitem[\protect\citeauthoryear{Kumar et al.}{2008}]{2008MNRAS.388.1729K} Kumar P., Narayan R., Johnson J. L., et al., 2008, MNRAS, 388, 1729

\bibitem[\protect\citeauthoryear{Kumar \& Panaitescu}{2000}]{2000ApJ...541L..51K} Kumar P., Panaitescu A., 2000, \apjl, 541, 51

\bibitem[\protect\citeauthoryear{Lang et al.}{2010}]{2010AJ...15164.1T} Lang D., Hogg D.~J., Mierle, K. et al., 2010, AJ, 139, 1782

\bibitem[\protect\citeauthoryear{Laskar et al.}{2013}]{2013GCN...15162...1L} Laskar T., Zauderer A., Berger E., 2013, GCN Circ. 15162

\bibitem[\protect\citeauthoryear{Liang et al.}{2006}]{2006ApJ...646..351L} Liang E.-W.,  Zhang B., O'Brien, P., et al., 2006, \apj, 646, 351

\bibitem[\protect\citeauthoryear{Liang et al.}{2007}]{2007APJ...670..565L} Liang E.-W.,  Zhang B.-B., Zhang B., 2007, \apj, 670, 565

\bibitem[\protect\citeauthoryear{L\"u \& Zhang}{2014}]{luz13} L\"u, H.~J., Zhang, B. 2014, \apj, 785, 74

\bibitem[\protect\citeauthoryear{L\"u et al.}{2015}]{2015ApJ...805...89L} L\"u, H.~J., Zhang, B., Lei, W.-H., \apj, 805, 89

\bibitem[\protect\citeauthoryear{Lyons et al.}{2010}]{2013MNRAS.402..705L} Lyons N., O'Brien P.~T., Zhang B., et al., 2010, MNRAS, 402, 705

\bibitem[\protect\citeauthoryear{Margutti et al.}{2011}]{2011MNRAS.410.1064M} Margutti R., Bernardini G., Barniol-Duran R., et al.,  2011, MNRAS, 410, 1064

\bibitem[\protect\citeauthoryear{Matheson et al.}{2003}]{2003ApJ...599..394M} Matheson T., Garnavich P. M., Stanek K. Z., et al., 2003, \apj, 599, 394

\bibitem[\protect\citeauthoryear{Metzger et al.}{2011}]{2011MNRAS.413.2031M} Metzger B.~D., Giannios D., Thompson T.~A., et al., 2011, MNRAS, 413, 2031

\bibitem[\protect\citeauthoryear{Metzger et al.}{2015}]{2015arXiv150802712M} Metzger B.~D.,  Margalit, B., Kasen, D., et al., 2015, MNRAS Letters submitted, arXiv:1508.02712

\bibitem[\protect\citeauthoryear{Molinari et al.}{2011}]{2007AA...469L..13M} Molinari E., Vergani S., Malesani D., et al., 2007, A\&A, 469, 13

\bibitem[\protect\citeauthoryear{Molotov et al.}{2008}]{2008AdSpR..41.1022M } Molotov I., Agapov V., Titenko V. D., et al., 2008, AdSpR, 41, 1022

\bibitem[\protect\citeauthoryear{Nousek et al.}{2006}]{2006ApJ...642..389N} Nousek J.~A., Kouveliotou C., Grupe D., et al., 2006 \apj, 1, 389

\bibitem[\protect\citeauthoryear{Oates et al.}{2009}]{2009MNRAS.395..490O} Oates S.~R., Page M.~J., Schady P., et al., 2009, MNRAS, 395, 490

\bibitem[\protect\citeauthoryear{Panaitescu \& Kumar}{2002}]{2002ApJ...571..779P} Panaitescu A. \& Kumar P., 2002, \apj, 571, 779

\bibitem[\protect\citeauthoryear{Pei}{1992}]{pei92} Pei Y.-C., 1992, \apj, 395, 130

\bibitem[\protect\citeauthoryear{Perley et al.}{2014}]{2014ApJ...781..37P} Perley D. A., Cenko S. B., Corsi A., et al., 2014, \apj, 781, 37

\bibitem[\protect\citeauthoryear{Poole et al.}{2008}]{2008MNRAS.383..627P} Poole T.~S., Breeveld A.~A., Page M.~J., et al., 2008, MNRAS, 383, 627

\bibitem[\protect\citeauthoryear{Pozanenko et al.}{2013a}]{2013GCN...15190...1P} Pozanenko A., Volnova A., Hafizov B., et al., 2013a, GCN Circ. 15190

\bibitem[\protect\citeauthoryear{Pozanenko et al.}{2013b}]{2013GCN...15237...1P} Pozanenko A., Volnova A., Hafizov B., et al., 2013b, GCN Circ. 15237

\bibitem[\protect\citeauthoryear{Pozanenko et al.}{2013c}]{2013EAS....61..259P} Pozanenko A., Elenin L., Litvinenko E., et al., 2013c, EAS  Pub. Series, 61, 259

\bibitem[\protect\citeauthoryear{Racusin et al.}{2009}]{2009ApJ...698...43R} Racusin J.~L., Liang E.~W., Burrows D.~N., et al., 2009, \apj, 698, 43

\bibitem[\protect\citeauthoryear{Reichart et al.}{2005}]{2005NCimC..28..767R} Reichart D., Nysewander M., Moran J., et al., 2005, NCimC, 28, 767

\bibitem[\protect\citeauthoryear{Rieger et al.}{2007}]{2007Ap&SS.309..119R} Rieger F., Bosch-Ramon F., Duffy P., 2007, Ap\&SS, 309, 119

\bibitem[\protect\citeauthoryear{Rowlinson et al.}{2013}]{2013MNRAS.430.1061} Rowlinson A., O'Brien P.~T., Metzger B.~D., et al., 2013, MNRAS, 430, 1061

\bibitem[\protect\citeauthoryear{Roming et al.}{2005}]{2005SSRv..671...95R} Roming P.~W.~A., Kennedy T.~E., Mason K.~O., et al., 2005, SSRV, 120, 95

\bibitem[\protect\citeauthoryear{Sari et al.}{1999}]{1999ApJ...519L..17S} Sari R., Piran T., Halpern, J. P., 1999, \apjl, 519, 17

\bibitem[\protect\citeauthoryear{Swenson et al.}{2013}]{2013ApJ...774....2S} Swenson C.~A., Roming P.~W.~A. De Pasquale, M., et al., 2013 \apj, 774, 2

\bibitem[\protect\citeauthoryear{Schady et al.}{2010}]{2010MNRAS.401.2773S} Schady P., Page M.~J., Oates S.~R., et al., 2010, MNRAS, 401, 2773

\bibitem[\protect\citeauthoryear{Schlegel et al.}{1998}]{1998ApJ...500..525S} Schlegel D.~J., Finkbeiner D.~P., Davis, M., 1998, \apj, 500, 525

\bibitem[\protect\citeauthoryear{Shen \& Zhang}{2009}]{shz09} Shen R.-F. \& Zhang B. M\'esz\'aros P., 2009, MNRAS, 398, 1936

\bibitem[\protect\citeauthoryear{Thompson}{2007}]{2007RMxAC..27...80T} Thomson T. A., 2007, RMxAC, 27, 80

\bibitem[\protect\citeauthoryear{Thompson et al.}{2010}]{ 2010AIPC.1279...81T} Thomson T. A., Metzger B. D., Bucciantini, N., 2010, AIPC, 1279, 81

\bibitem[\protect\citeauthoryear{Troja et al.}{2007}]{2007ApJ...665..599T} Troja E., Cusumano, G., O'Brien P.~T., et al., 2007 \apj, 665, 599

\bibitem[\protect\citeauthoryear{Tody}{1986}]{1986SPIE..627..733T} Tody D., 1986, Society of Photo-Optical Instrumentation Engineers (SPIE) Conference Series, 627, 733
 
\bibitem[\protect\citeauthoryear{Tody}{1993}]{tody1993} Tody D., 1993  Astronomical Data Analysis Software and Systems II, A.S.P. Conference Ser., Vol 52, eds. R.J. Hanisch, R.J.V. Brissenden, \& J. Barnes, 173 

\bibitem[\protect\citeauthoryear{Trotter et al.}{2013}]{2013GCN...15164.1T} Trotter A., Haislip J., Lacluyze A., et al., 2013, GCN Circ. 15148
 
\bibitem[\protect\citeauthoryear{Usov}{1992}]{1992Natur.357..472U} Usov V.~V., Nature, 357, 472

\bibitem[\protect\citeauthoryear{Van Eerten et al.}{2010}]{2007ApJ...722..235} van Eerten H., Zhang, W., MacFadyen, A., 2010, \apj, 722, 235

\bibitem[\protect\citeauthoryear{Vietri \& Stella}{1998}]{1998ApJ...507L..45V} Vietri M., Stella, L. 1998, \apjl, 507, 45

\bibitem[\protect\citeauthoryear{Volnova et al.}{2013a}]{volnova15185}  Volnova A., Matkin A., Stepura A., et al., 2013a, GCN Circ. 15185

\bibitem[\protect\citeauthoryear{Volnova et al.}{2013b}]{volnova15188}  Volnova A.,  Linkov V.,  Polyakov K.,  et. al., 2013b, GCN Circ. 15188

\bibitem[\protect\citeauthoryear{Volnova et al.}{2013c}]{volnova15189}  Volnova A.,  Krugly Y.,  Slyusarev I.,  et. al., 2013c, GCN Circ. 15189

\bibitem[\protect\citeauthoryear{Watson et al.}{2012}]{2012SPIE.8444.5LW} Watson A.~M., Richer M.~G., Bloom J.~S., et al., 2012, SPIE, 8444, 5

%\bibitem[Cano (2011)]{2011MNRAS.413..699C} Cano, Z., Bersier, D., Guidorzi C., et al., 2011, MNRAS, 413, 669

\bibitem[\protect\citeauthoryear{Woosley}{2011}]{wosley2011} Woosley S.E., 2011, in "Gamma-Ray Bursts", Cambridge University Press, http://arxiv.org/abs/1105.4193

\bibitem[\protect\citeauthoryear{Zauderer}{2013}]{Zauderer2013} Zauderer B. A., Laskar T., \& Berger E., 2013, GCN Circ. 15159

\bibitem[\protect\citeauthoryear{Zhang et al.}{2006}]{2006ApJ...642..354Z} Zhang B., Fan Y.-Z., Dyks J., et al., 2006, \apj, 642, 345

\bibitem[\protect\citeauthoryear{Zhang}{2011}]{2011CRPhy..12..206Z} Zhang B., 2011, CRPhy, 12, 206

\bibitem[\protect\citeauthoryear{Zhang et al.}{2007}]{2007ApJ...655..989Z} Zhang B., Liang E.-W., Page K., et al., 2007, \apj, 655, 989

\bibitem[\protect\citeauthoryear{Zhang \& MacFadyen}{2009}]{2009ApJ...698.1261Z} Zhang W.-Q., MacFadyen, A., 2009, \apj, 698, 1261

\bibitem[\protect\citeauthoryear{Zhang \& M\'esz\'aros}{2001}]{zms01} Zhang B. \& M\'esz\'aros P., 2001, \apjl, 552, 35

\bibitem[\protect\citeauthoryear{Zhang \& M\'esz\'aros}{2002}]{zms02} Zhang B. \& M\'esz\'aros P., 2002, \apj, 566, 712

\bibitem[\protect\citeauthoryear{Zhang}{2009}]{zhang2011} Zhang X.-H., 2009, Research in Astronomy and Astrophysics, 9, 213

%\bibitem[Auri\`ere(1982)]{aur82} Auri\`ere, M.  1982, \aap,
%    109, 301
%\bibitem[Canizares et al.(1978)]{can78} Canizares, C. R.,
%    Grindlay, J. E., Hiltner, W. A., Liller, W., \&
%    McClintock, J. E.  1978, \apj, 224, 39
%\bibitem[Djorgovski \& King(1984)]{djo84} Djorgovski, S.,
%    \& King, I. R.  1984, \apjl, 277, L49
%\bibitem[Hagiwara \& Zeppenfeld(1986)]{hag86} Hagiwara, K., \&
%    Zeppenfeld, D.  1986, Nucl.Phys., 274, 1
%\bibitem[Harris \& van den Bergh(1984)]{har84} Harris, W. E.,
%    \& van den Bergh, S.  1984, \aj, 89, 1816
%\bibitem[H\`enon(1961)]{hen61} H\'enon, M.  1961, Ann.d'Ap., 24, 369
%\bibitem[Heiles \& Troland(2003)]{heiles03} Heiles, C. \& Troland, T. H., 2003, \apjs, preprint doi:10.1086/381753
%\bibitem[Kim, Ostricker, \& Stone(2003)]{kim03} Kim, W.-T.,  Ostriker, E., \& Stone, J. M., 2003, \apj, 599, 1157
%\bibitem[King(1966)]{kin66}  King, I. R.  1966, \aj, 71, 276
%\bibitem[King(1975)]{kin75}  King, I. R.  1975, Dynamics of
%    Stellar Systems, A. Hayli, Dordrecht: Reidel, 1975, 99
%\bibitem[King et al.(1968)]{kin68}  King, I. R., Hedemann, E.,
%    Hodge, S. M., \& White, R. E.  1968, \aj, 73, 456
%\bibitem[Kron et al.(1984)]{kro84} Kron, G. E., Hewitt, A. V.,
%    \& Wasserman, L. H.  1984, \pasp, 96, 198
%\bibitem[Lynden-Bell \& Wood(1968)]{lyn68} Lynden-Bell, D.,
%    \& Wood, R.  1968, \mnras, 138, 495
%\bibitem[Newell \& O'Neil(1978)]{new78} Newell, E. B.,
%    \& O'Neil, E. J.  1978, \apjs, 37, 27
%\bibitem[Ortolani et al.(1985)]{ort85} Ortolani, S., Rosino, L.,
%    \& Sandage, A.  1985, \aj, 90, 473
%\bibitem[Peterson(1976)]{pet76} Peterson, C. J.  1976, \aj, 81, 617
%\bibitem[Rudnick et al.(2003)]{rudnick03} Rudnick, G. et al., 2003, \apj, 599, 847
%\bibitem[Spitzer(1985)]{spi85} Spitzer, L.  1985, Dynamics of
%    Star Clusters, J. Goodman \& P. Hut, Dordrecht: Reidel, 109
%\bibitem[Treu et al.(2003)]{treu03} Treu, T. et al., 2003, \apj, 591, 53
\end{thebibliography}
\end{document}